\documentclass[useAMS,usenatbib,onecolumn]{mnras}
\usepackage{amsmath,bm,amssymb}
\usepackage{dcolumn}
\usepackage{graphics,graphicx}
\usepackage{float}
\usepackage{threeparttable}
\usepackage{url}
\usepackage{epstopdf}

\usepackage{multirow}

\usepackage{color}

\usepackage{longtable}
\usepackage{url}

\title[ExoMol line lists -- LII. CH$^{+}$]{ExoMol line lists -- LII.  Line Lists for the Methylidyne Cation (CH$^{+}$)}

\author[Pearce et al.]
{Oliver Pearce,  Sergei N. Yurchenko and Jonathan Tennyson\thanks{Corresponding author: j.tennyson@ucl.ac.uk}\vspace*{4mm}\\
Department of Physics and Astronomy, University College London, Gower Street, WC1E 6BT London, UK}

\date{Accepted XXXX. Received XXXX; in original form XXXX}
\pagerange{\pageref{firstpage}--\pageref{lastpage}} \pubyear{2023}
\begin{document}
\label{firstpage}
\maketitle

\begin{abstract}
Comprehensive and accurate rovibronic line lists for the
X\,$^{1}\Sigma^{+}$ and A\,$^{1}\Pi$ states of $^{12}$C$^{1}$H$^{+}$ and $^{13}$C$^{1}$H$^{+}$ which should be
applicable up to temperatures of 5000 K are presented.
Available empirical potential energy curves and high-level \textit{ab initio} dipole and
transition dipole moment curves are used with the
program LEVEL to compute  rovibronic energy levels and  Einstein $A$ coefficients. $\Lambda$-doubling is incorporated into the energy
levels and $A$-coefficients involving the A\,$^{1}\Pi$ state
using an empirical method. For $^{12}$C$^{1}$H$^{+}$,  line positions are improved by using
both laboratory and astronomical
observational spectra as input to the MARVEL procedure.
The  $^{12}$C$^{1}$H$^{+}$ line list contains 1505 states and 34\,194 transitions over the
frequency range of 0 -- 33\,010 cm$^{-1}$ ($\lambda > 300$ nm).  Comparisons with
observed astronomical and laboratory spectra give  very good agreement.
The {PYT} CH$^{+}$ line lists and partition functions are available from the ExoMol database at \href{http://www.exomol.com}{www.exomol.com}
\end{abstract}


\section{Introduction}
The methylidyne cation, CH$^{+}$, was one of the first molecules, and the
first molecular ion, to be detected in  space \citep{37Dunham}.
CH$^+$ has since  been found to be ubiquitous in  interstellar space \citep{23GoPiHe.CH+} where it has been
observed both in absorption and emission in various cool
environments  \citep{21NeGoChFa,20KeGaBo.CH+,13YoDaWe.CH+,10NaDaHa.CH+,97CeLiGoCo,93GrVaBl.CH+,89Crawford.CH+,37Dunham} which include protostellar
disks \citep{11ThMeMe.CH+}, diffuse clouds \citep{95CrLaSh.CH+} and
lines of sight towards star-forming regions
\citep{10FaGoCe.CH+}, with even
routine detection from extra-galactic sources
\citep{17FaZwGoBe.CH+,17MuMuBlGe.CH+,11RaNaMaPh.CH+,15RoWeDaYo.CH+}. In addition,
strong CH$^+$ A\,$^{1}\Pi$ -- X\,$^{1}\Sigma^{+}$  emission spectra have been observed in the red rectangle (post–asymptotic giant branch
 star HD 44179), see \citet{04HoThOk.CH+}  and references therein.

Astrophysical interest in  CH$^+$ has sparked many
investigations; these include using its unique properties, which allow  tracing of
significant energy releases and interstellar turbulence
\citep{22ViFaBaGo}, and its importance in astrochemistry,
particularly for helping to understand diffuse interstellar clouds.
CH$^{+}$ is thought to be a building block for the formation of organic
molecules in interstellar space, principally through its role as an intermediary
in reactions producing larger hydrocarbons \citep{04BaVa.CH+,
18DoJuScAs}, and molecules including C$_{2}$, CN, CO and CH
\citep{07HaKeSzZa}. Furthermore, owing to its efficient destruction
mechanisms \citep{93GrVaBl.CH+} and inherently reactive nature with
species prevalent in its environment (most notably H and H$_{2}$), it came as
a great surprise when the abundance of the molecule was observed to exceed
predictions by several orders of magnitude \citep{76Dalgarno.CH+}. This has stimulated
significant investigation into the formation, destruction and abundance
of CH$^{+}$ over many years \citep{93GrVaBl.CH+,17VaGoHe.CH+,10FaGoCe.CH+,97Gredel.CH+,10WeCeBa.CH+,97Gredel.CH+}.
While the
primary formation mechanism and observed abundance of CH$^{+}$ remain in
question, some recent studies have begun to provide
explanations that reduce the large discrepancy between calculation and
observation \citep{jt688,23GoPiHe.CH+}.

Interest in CH$^{+}$ has also motivated a number of laboratory studies. Its
rovibronic spectrum (involving vibrationally and rotationally-resolved transitions between electronic states) has been
widely studied
\citep{07HeRoLa.CH+,06HaKeSz.CH+,04DuLe,
02HeWiLaLi,86CaSo.CH+,86SaWaWh,82CaRa,
81GrBrOkWi,82HeCoGrMo,80CoHeMo,60DoMo,
42DoHe.CH+,73BoLoVe,10Mueller,18YuDrPe.CH+}, most
notably the A\,$^{1}\Pi$ - X\,$^{1}\Sigma^{+}$ (A-X) system which covers
transitions between the two lowest energy singlet states.
Landmark studies include the work by
\citet{82CaRa}, the first high-resolution study (uncertainty $\leq
0.01$ cm$^{-1}$) which also assigned many transitions involving low
vibrational levels ($v'' = 0-3$), and more recently that of
\citet{06HaKeSz.CH+} who comprehensively characterised low vibrational
bands. The pure rotational spectrum of CH$^{+}$, first detected in
space through emission
lines from a planetary nebula \citep{97CeLiGoCo}, has received much less
coverage with only a few low-energy lines being observed so far. The first
laboratory detection of a pure rotational line ($J=1-0$)
\citep{06PeDr} was later proven to be inaccurate by 0.0019
cm$^{-1}$ and its frequency was superseded by later studies
\citep{18DoJuScAs,10Amano}.
The rovibrational spectrum has proven more elusive still, the first and only
high resolution laboratory study was made by \citet{18DoJuScAs}, and the corresponding
astronomical detection only from one source, NGC 7027
\citep{21NeGoChFa}. The minor isotopologues, most notably
$^{13}$CH$^{+}$ and $^{12}$CD$^{+}$, have also received significant
spectroscopic investigation
\citep{18DoJuScAs,87BeCiKe,10Amano,04DuLe,15AmPeDrYu, 07HeRoLa.CH+,18YuDrPe.CH+,97Bembenek}.
A summary of the spectroscopic laboratory and
astronomically observed literature used in this study that covers the relevant
electronic states, X\,$^{1}\Sigma^{+}$ and A\,$^{1}\Pi$, is given in Table~\ref{tab:observation}.
Some papers provide spectroscopic analysis but do not present
new line data
\citep{10Amanob,18YuDrPe.CH+,15Amano,10Mueller,07HaKeSzZa}.

\begin{figure}
    \centering
    \includegraphics[width=0.5\textwidth]{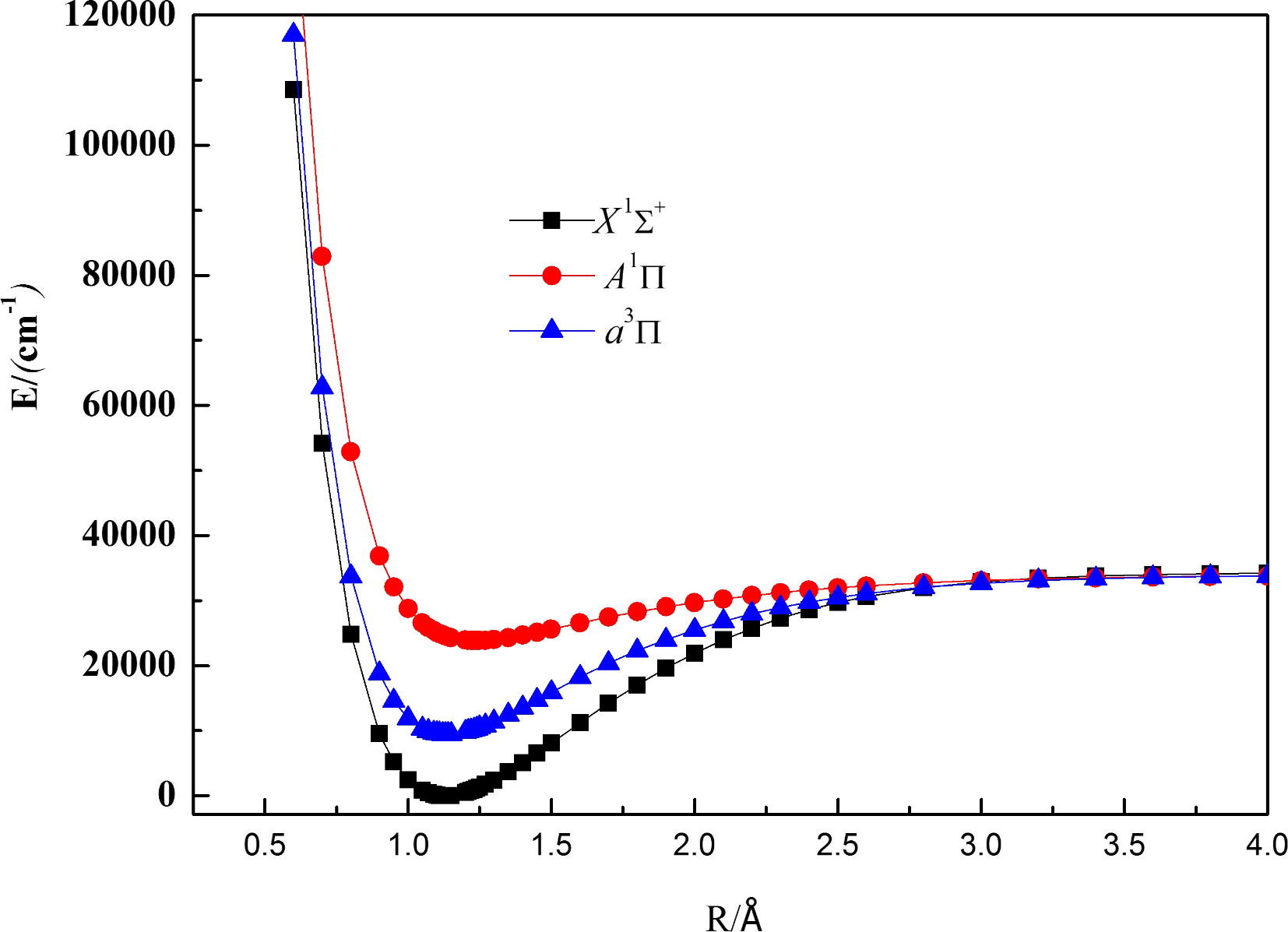}
    \caption{Potential energy curves for the low-lying electronic states below
the first dissociation limit, \protect{X\,$^{1}\Sigma^{+}$, A\,$^{1}\Pi$ and
a\,$^{3}\Pi$}. Figure from \citet{17GaWuWa.CH+}.}
    \label{fig:PECs}
\end{figure}

\begin{figure}
    \centering
        \includegraphics[width=.5\textwidth]{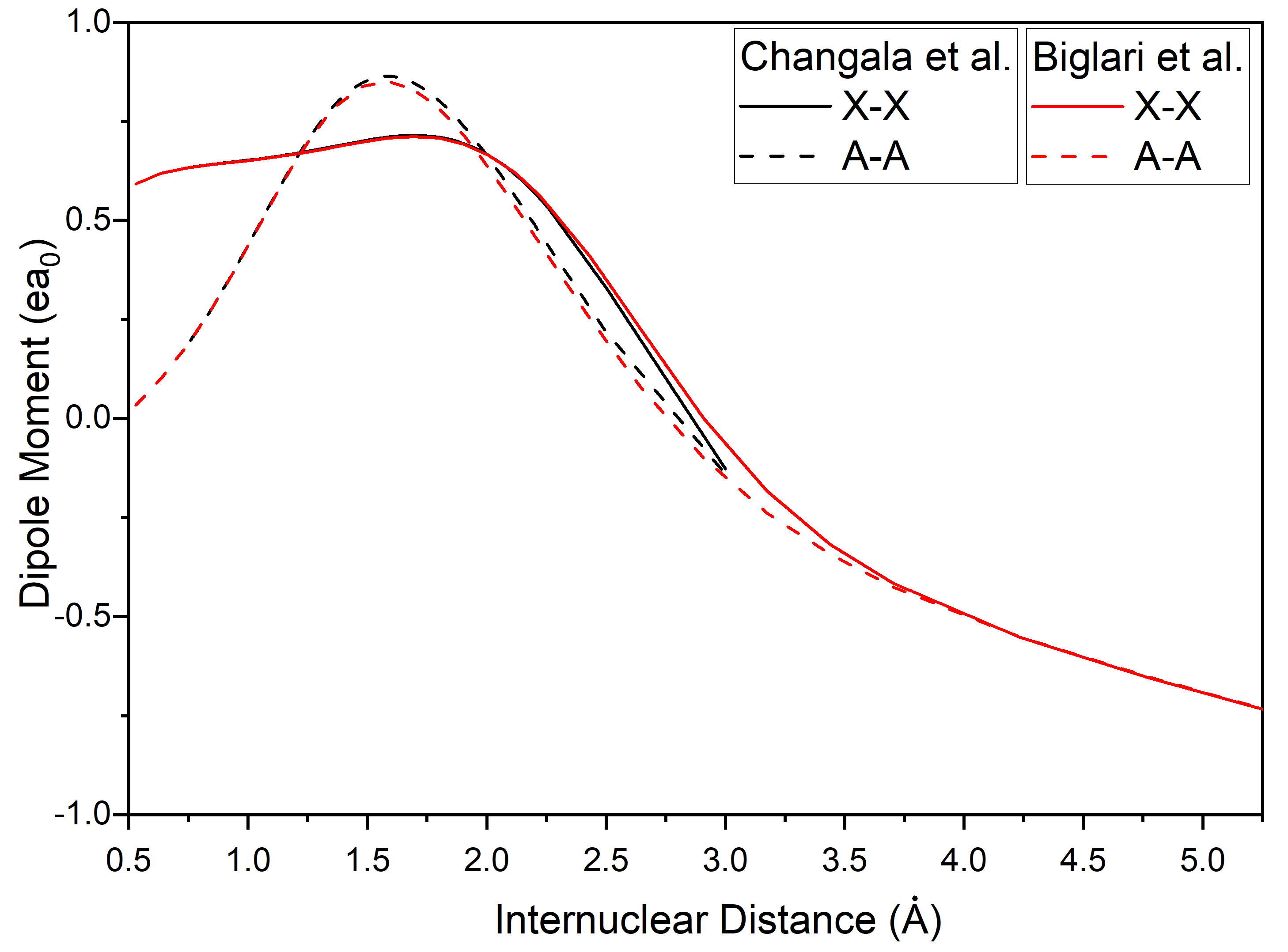}
        \includegraphics[width=.5\textwidth]{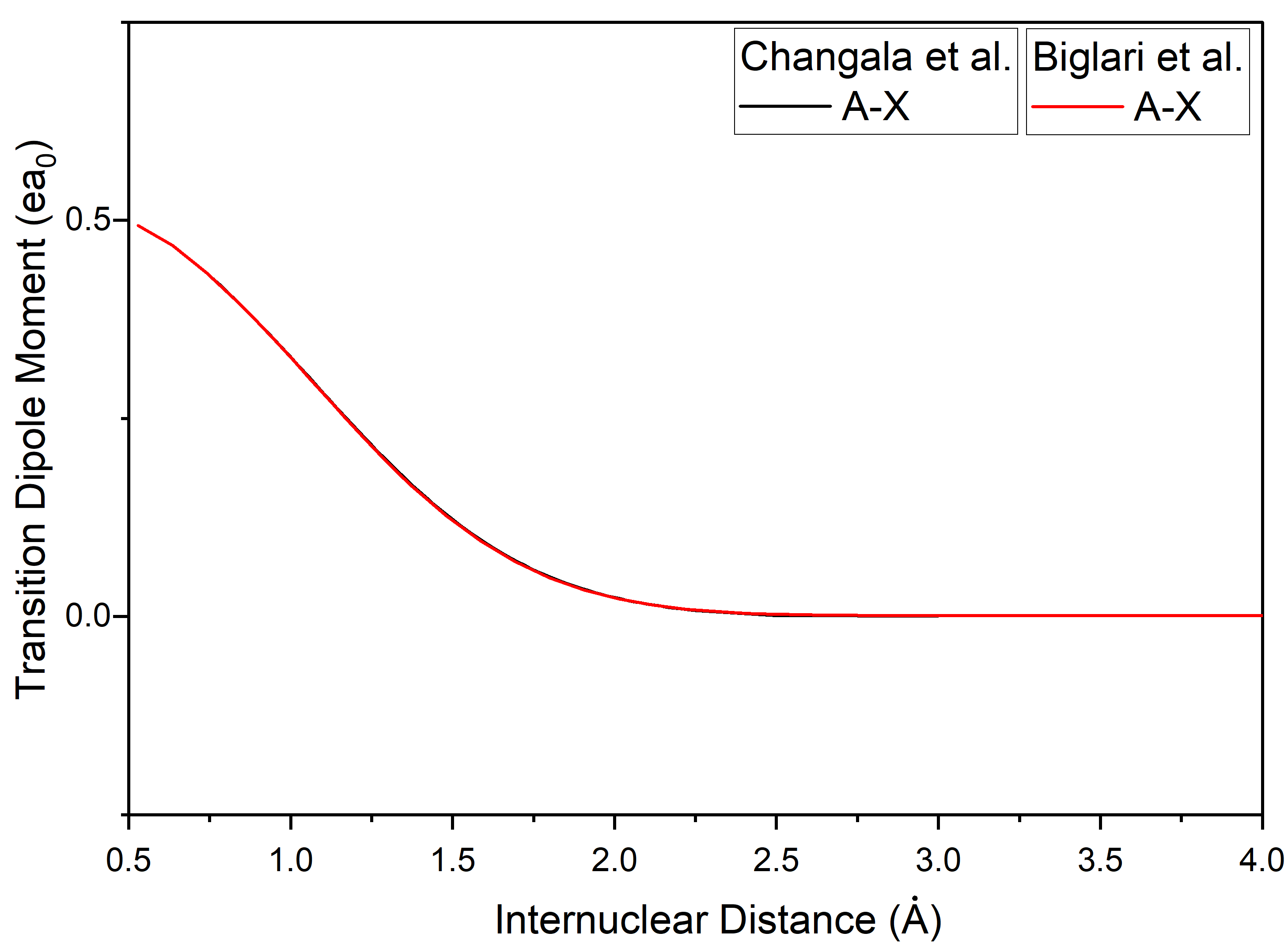}
    \caption{Comparison of dipole moments, in a.u., for the \protect{X\,$^{1}\Sigma^{+}$ and
A\,$^{1}\Pi$ states (Top) and transition dipole moments for A\,$^{1}\Pi$ -
X\,$^{1}\Sigma^{+}$} (Bottom) from \citet{14BiShMa.CH+} and  \citet{21ChNeGo.CH+}.}
    \label{fig:dms}
\end{figure}

Theoretical investigations have aimed to characterise the potential
energy curves (PECs) and dipole/transition dipole moments (DMs/TDMs) of
CH$^{+}$. PECs, unique to each electronic state, describe the potential energy
of an electron as a function of nuclear distance, while DMs/TDMs describe the
strength of interaction between two different states during a transition between
them. Many \textit{ab initio} electronic structure studies have
focused on CH$^{+}$ since it
contains just six electrons; these provide  PECs for
low-lying electronic states, notably X\,$^{1}\Sigma^{+}$, A\,$^{1}\Pi$ and
a\,$^{3}\Pi$ \citep{21ChNeGo.CH+,17GaWuWa.CH+,13SaSp.CH+}, with some also
characterising higher energy states
\citep{23ToNkOw.CH+,17BaMc.CH+,14BiShMa.CH+,04BaVa.CH+,01KoPi.CH+,91KaSuFr.CH+,
80SaKiLi.CH+}. Here we use the PECs due to  \citet{16ChLe.CH+} who used
a fully empirical (`direct potential fit') approach that utilises
experimental data to fit the PECs of the X\,$^{1}\Sigma^{+}$ and
A\,$^{1}\Pi$ states which, if sufficient experimental data is available, is a more accurate method than \textit{ab initio} studies.
The PECs of the three lowest electronic states of CH$^{+}$ can be seen in Fig.~\ref{fig:PECs} \citep{17GaWuWa.CH+}.
Several  studies
have characterised DMs/TDMs
\citep{21ChNeGo.CH+,17GaWuWa.CH+,17BaMc.CH+,13SaSp.CH+,
91KaSuFr.CH+,80SaKiLi.CH+} with notable investigations covering a vast number of
singlet, triplet and quintet states using large basis sets and extending to
large internuclear separations \citep{23ToNkOw.CH+,14BiShMa.CH+,jt695}. The curves were
compared (where data was available) and were seen to agree well, particularly
those of A-X. A comparison of the X-X and A-A DMs and A-X TDMs from
\citet{14BiShMa.CH+} and  \citet{21ChNeGo.CH+} is shown in
Fig.~\ref{fig:dms}.

The ExoMol project aims to provide comprehensive spectroscopic data for
transitions within molecules relevant to exoplanetary atmospheres, cool stars
and brown dwarfs \citep{jt528}. Since its inception, the project has created
line lists for many diatomic molecules, ions and larger polyatomic
species, that comprehensively
characterise the spectroscopic properties of the molecule and allows a
model/profile to be built up.
In this study, we calculate a comprehensive and accurate line
list for the  X\,$^{1}\Sigma^{+}$ and A\,$^{1}\Pi$ states of
$^{12}$C$^{1}$H$^{+}$ and $^{13}$C$^{1}$H$^{+}$  as part of the ExoMol project.

\begin{table}
\centering
\caption{Summary of spectroscopic laboratory and astronomically observed
papers used in our MARVEL study. } \label{tab:observation}
{\small\tt
\begin{tabular}{lclcl}
\hline\hline
\multicolumn{1}{l}{Reference} & \multicolumn{1}{c}{Freq. Range (cm$^{-1}$)}  & \multicolumn{1}{l}{Detected Transitions\textit{$^{a}$}} & \multicolumn{1}{c}{A/V\textit{$^{b}$}} & \multicolumn{1}{l}{Uncert\textit{$^{c}$}  (cm$^{-1}$)}     \\ \hline
21NeGoChFa \citep{21NeGoChFa} & 2422.62 - 2942.57  & $(1,0)$: $J''\leq10$  &16/16 & $3\times 10^{-2}$  \\
18DoJuScAs \citep{18DoJuScAs} & \begin{tabular}[c]{@{}c@{}} 27.85719\\ 2711.81 - 2817.23\end{tabular} & \begin{tabular}[c]{@{}l@{}} $(0,0)$: R(0) \\ $(1,0)$: $J''\leq3$ \end{tabular} & 5/5 & $2\times 10^{-5}$  \\
15YuDrPeAm \citep{15AmPeDrYu} & 55.68 - 83.44 &  $(0,0)$: R(1), R(2) & 2/2 & $3\times 10^{-6}$  \\
10Amano \citep{10Amano} & 27.85719 & $(0,0)$: R(0) & 1/1 & $7\times 10^{-7}$ \\
06PeDr \citep{06PeDr} & 27.85523  &$(0,0)$: R(0)& 1/0 & $3\times 10^{-6}$  \\
06HaKeSzZa \citep{06HaKeSz.CH+} & 20401.56 - 23967.41 & $(0,0),(0,1),(2,1)$: $J''\leq17$& 124/124 & $1\times 10^{-2}$  \\
02HeWiLaLi \citep{02HeWiLaLi} & 31747.0 - 33000.0  & \begin{tabular}[t]{@{}l@{}} $(11,0),(12,0),(13,0),(14,0): $ \\$J''\leq9$ \end{tabular} & 35/35 & 1  \\
97CeLiGoCo \citep{97CeLiGoCo} & 55.68 - 165.92  &  $(0,0): J''=1-5$  & 6/6 & $5\times 10^{-2}$  \\
89SaWhGr \citep{89SaWhGr} & 16750 - 18613  & $(0,1),(1,1),(2,1): J''=32-35$ & 6/6 & 1 \\
86SaWaWh \citep{86SaWaWh} & 16125.31 - 18613.07 &\begin{tabular}[t]{@{}l@{}}    $(0,0),(0,1),(1,1),(1,2),(2,1)$,\\$(2,2),(3,2),(3,3),(4,3),(5,3),$\\$(5,4),(7,4),(7,5): J''=18-35$ \end{tabular} & 29/29 & $5\times 10^{-3}$  \\
82CaRa \citep{82CaRa} & 17111.40 - 25281.30& \begin{tabular}[t]{@{}l@{}}     $(0,0),(0,1),(1,0),(1,1),(1,2),$\\$(1,3),(2,1),(3,1): J''\leq14$ \end{tabular} & 231/231 & $1\times 10^{-2}$  \\
82HeCoGrMo \citep{82HeCoGrMo} & 15460 - 28484 & \begin{tabular}[t]{@{}l@{}}Many $v'\leq10$ vibrational\\ bands: $J''=11-36$ \end{tabular}& 51/51 &  1  \\
81GrBrOkWi \citep{81GrBrOkWi} & 22540.0 - 23615.1 & $(0,0): J''=6-21$ & 36/36 & $1\times 10^{-1}$ \\
80CoHeMo \citep{80CoHeMo} & 27494 - 28523& Transitions unassigned & N/A & $1\times 10^{-2}$\\
60DoMo \citep{60DoMo} & 20524.30 - 26266.21  &\begin{tabular}[t]{@{}l@{}} $(0,1),(1,1),(2,1),(3,1),(4,1):$ \\ $J''\leq15$ \end{tabular}& 130/129 &$5\times 10^{-2}$ \\
42DoHe \citep{42DoHe.CH+} & 20636.8 - 26704.1  & $(0,0),(0,1),(1,0),(2,0): J''\leq11$ & 94/94 &$2\times 10^{-1}$  \\
\hline\hline
\end{tabular}
}
\mbox{}\\
{\flushleft
\textit{$^{a}$} Bracketed values indicate the vibrational bands involved in transition, ($v^\prime,v^{\prime\prime}$). \\
\textit{$^{b}$} `A/V': Number of actual lines in source/Number of lines validated  by MARVEL.\\
\textit{$^{c}$} Uncert: Average uncertainty as quoted (if available),
or the modified uncertainty for the MARVEL algorithm.
}
\end{table}

\section{Methods}
To calculate the line list of CH$^{+}$, quantum mechanical nuclear-motion calculations were
performed using the program LEVEL \citep{LEVEL} which generated the desired energy
levels and transition data.  To verify  and improve on
the calculated data, existing experimental and observational spectroscopic data
for CH$^{+}$ were gathered for use with the program MARVEL (Measured Active
Rotational-Vibrational Energy Levels) \citep{jt412}. The MARVEL \citep{jt412}  algorithm
inverts transition data to give
highly accurate energy levels along with determined uncertainties.

\subsection{MARVEL energy levels}
767 $^{12}$CH$^{+}$ lines were collected from existing observational studies  as summarised in Table~\ref{tab:observation}.
While most of the sources are from high resolution laboratory studies, 97CeLiGoCo \citep{97CeLiGoCo} and
21NeGoChFa \citep{21NeGoChFa} are actually astronomical observations.
Table~\ref{tab:observation} shows that there are rather few observed
transitions within the X\,$^{1}\Sigma^{+}$ ground  state of   $^{12}$CH$^{+}$, hence the need to
include astronomically observed lines which were deemed sufficiently accurate to use and helped to expand
and refine the MARVEL  network. The vast majority of existing data
covers the A-X band with variable accuracy, with some papers, particularly older ones, not providing
uncertainties at all and, in the case of 80CoHeMo \citep{80CoHeMo},  not assigning their transitions. This means that the observed frequencies cannot be added to the MARVEL network, where the nodes are characterised by the quantum numbers of the energy levels involved. There is
only comprehensive MARVEL energy level data (extending up to high $J$) for vibrational states with
$v=0-3/4$ for both electronic states, and some limited data for A\,$^{1}\Pi$ $v=11-14$.

MARVEL analyses uncertainty when evaluating transition data and is able to
determine whether or not an uncertainty is consistent with the other data.
Using this, the quoted experimental uncertainties for the older sources 42DoHe \citep{42DoHe.CH+}, 60DoMo \citep{60DoMo} and
81GrBrOkWi \citep{81GrBrOkWi}  were increased to allow MARVEL to become self-consistent.
After this it was only necessary to exclude two lines from the network: the  06PeDr pure rotational line and the 60DoMo A-X $v=2-1$ P(8) line.

The MARVEL input transitions file and output energies file are given as supplementary data to this paper. There was insufficient
spectroscopic data to perform a MARVEL study for $^{13}$CH$^{+}$.

\subsection{Calculations (LEVEL)}
Version LEVEL-16 \citep{LEVEL} was used to compute the line list. This solves the radial Schr\"{o}dinger equation to give vibrational and rotational energy levels within a defined potential, as well as transition frequencies and Einstein coefficients for the coupling within or between potentials. LEVEL was chosen as the existing work on CH$^{+}$ by \citet{16ChLe.CH+} used features in this program that are not easy to reproduce with
other nuclear motion programs.

The effective radial Schr\"{o}dinger equation used in this study as given by \citet{16ChLe.CH+} is
\begin{equation}\label{eq:se}
    \left \{ -\frac{\hbar^2}{2\mu}\frac{d^2}{dr^2} + V_i(r) + \frac{\left [ J(J+1)-\Lambda^2\right]\hbar^2}{2\mu r^2}\left [ 1+g_i(r) \right ]  \right \}\psi_{v,J}(r)=E_{v,J}\, \psi_{v,J}(r)
\end{equation}
where $\mu$ is the reduced mass and $\Lambda$ represents the projection of electronic orbital angular momentum on the internuclear axis. $V_{i}(r)$ is the adiabatic PEC of electronic state $i$ that can include Born-Oppenheimer breakdown (BOB) functions and $g_{i}(r)$ describes centrifugal BOB function; see \citet{16ChLe.CH+} for more detail on these BOB functions.
For the ($\Lambda=1$) A\,$^{1}\Pi$ state, Cho \& Le Roy also include a
$J$-dependent term in the potential to incorporate the effects of
$\Lambda$-doubling, as follows
\begin{equation}\label{eq:l-doubling}
    sg(e/f)\Delta V_{\Lambda}(r) \left [ J(J+1) \right ]^\Lambda = sg(e/f)\left
\{ \left ( \frac{\hbar^2}{2\mu r^2} \right )^{2\Lambda}f_{\Lambda}(r)\right
\}\left [ J(J+1) \right ]^\Lambda
\end{equation}
where $sg(e/f) =+1$ and 0 for $e$ and $f$ levels respectively (in
accordance with the form suggested later by  \citet{18YuDrPe.CH+}),
and $f_{\Lambda}(r)$ is a $\Lambda$-doubling radial strength function with
the following form, also determined by Cho \& Le Roy
\begin{equation}\
    f_{\Lambda}(r)= \sum_{i=0}^{4}\omega _{i}y_{p_{\Lambda}}^{i}
\end{equation}
with
\begin{equation}
y_{p_{\Lambda}}(r)=\frac{r^{p_{\Lambda}}-r_{e}^{p_{\Lambda}}}{r^{p_{\Lambda}}+r_
{e}^{p_{\Lambda}}}
\end{equation}
where $\omega_{i}$ are expansion coefficients, $r_{e}=1.235896$ \AA\:is the
equilibrium internuclear distance for the A\,$^{1}\Pi$ state potential, and
$p_{\Lambda}=4$, all determined and given in  \citet{16ChLe.CH+}.
Solving Eq.~(\ref{eq:se}) for each of the two electronic state potentials,
$V_{i}(r)$, yields the vibrational and rotational eigenfunctions (energy
levels) that form the basis of the line list.

\subsubsection{Potential Energy and Dipole Moment Curves}
The CH$^{+}$ input file created for use in LEVEL by \citet{16ChLe.CH+}
was used as a basis in this study.  Our investigation was limited to the ground and
first excited singlet states, X\,$^{1}\Sigma^{+}$ and A\,$^{1}\Pi$, as
investigated by Cho \& Le Roy. As shown in Fig.~\ref{fig:PECs}
these are the only singlet states below
the first dissociation limit, and any spin-forbidden transitions between these states
and the low-lying triplet (a\,$^{3}\Pi$) state are yet to be observed and likely to be extremely weak  \citep{17GaWuWa.CH+}.
Spin-orbit coupling to this state was also neglected.

The full, empirical potential energy curves calculated by \citet{16ChLe.CH+}
utilise previous spectroscopic data to characterise the curves
more accurately than previous \textit{ab initio} studies. Literature DMs (X-X and A-A) and
TDMs (A-X) were compared to determined the most suitable
available dipole moment curves, see Fig.~\ref{fig:dms}. Since all curves agree well, we chose those
 of \citet{14BiShMa.CH+},
due to their comprehensive characterisation of the
curves up to high internuclear distance, $r$. We note   that \citet{23ToNkOw.CH+} also recently characterised
these quantities, but their work only became available after we started calculations and
hence was not considered in the initial analysis.

To get correct results it transpired that the basic input file provided by
\citet{16ChLe.CH+} required modification.
We utilised the most recent version of LEVEL (LEVEL-16) but there were significant
issues with the both input file and the LEVEL code itself. Firstly, the input file
gave incorrect values for several parameters which had to be investigated and corrected.
For example, $\Lambda$ being set to 0 for the A\,$^{1}\Pi$ and  most of the parameters describing the
$^{12}$CH$^{+}$ X\,$^{1}\Sigma^{+}$ ground state were set to values that
describe the $^{12}$CD$^{+}$ A\,$^{1}\Pi$ excited state!  Even when
amended, the input file still gave errors when running with LEVEL.
Analysis of the Fortran source code showed that
in places the code disagreed with the documentation and input file structure on the
ordering of  parameters that read the input potential energy function; we therefore modified the
LEVEL code to resolve these errors. As \citet{16ChLe.CH+} did not consider transition probabilities,  we then
modified the input file
to read in the  DM/TDM functions and to
instruct LEVEL to compute Einstein A coefficients. The final version of the LEVEL input file is provided as supplementary info. The source for the  modified LEVEL can be found at \href{github.org/exomol}{github.org/exomol}.

\subsubsection{$\Lambda$-Doubling}

LEVEL does not consider $\Lambda$-doubling, the splitting of rotational energy levels in $\Lambda>0$
electronic states due to their $e/f$-parity. $\Lambda$-doubling is an important effect to consider, as without
it our calculated energy levels in the A\,$^{1}\Pi$ state would be degenerate which is incorrect.
\citet{18YuDrPe.CH+} analysed $\Lambda$-doubling in the CH$^+$
A\,$^{1}\Pi$ state where the effect arises due to the interactions of
the $e/f$ rotational levels with $^1\Sigma^{+}$/$^1\Sigma^{-}$ states,  respectively.
They found that  treatment of this effect in all previous spectroscopic studies was incorrect as it was generally assumed that
the $\Lambda$-doubling energy of $e/f$ levels would lead to a symmetric shift in their energy terms, as the magnitude  of the interaction of $e$ levels with $\Sigma^{+}$ states and $f$ levels with $\Sigma^{-}$ states was assumed to be equal. In practice, \citet{18YuDrPe.CH+}  argued that the energy splitting is asymmetric with the $f$ levels negligibly perturbed since $\Sigma^{-}$ states will lie at much higher energy (none have so far been detected).
In accordance with the findings of
\citet{18YuDrPe.CH+}, the A\,$^{1}\Pi$ state energies computed by LEVEL correspond to the energy of the unperturbed $f$-parity
level. It is therefore necessary to account for the
perturbed $e$-parity levels. \citet{16ChLe.CH+} model the splitting by the
inclusion of an additional term in the potential energy function of their radial
Schr\"{o}dinger equation, as described in Eq.~(\ref{eq:l-doubling}),
which shows that the $e$-parity energy level perturbation increases with
$J^{2}$, whereas the $f$-parity levels are unperturbed. Since the version of LEVEL we used does not
account for this, an empirical approach to considering the effect was
taken.

The approach taken consisted of separately
calculating a correction to each $f$-parity rotational level within each vibrational level of the A\,$^{1}\Pi$ state, which could be applied to give the energy of the corresponding $e$-parity level. This involved computing the expectation value of the operator $\Delta V_{\Lambda}(r)$ given in
Eq.~(\ref{eq:l-doubling}) for each vibrational level, $\left \langle
\psi_v |\Delta V_{\Lambda}(r)|\psi_v \right \rangle$, and evaluating the
rotational $J$-dependence to give the energy level correction. To do
this, the operator was calculated for each value of internuclear distance,
$r$, in steps of 0.001 \AA \:from $r_{\rm min}$ to $r_{\rm max}$ for each
vibrational level, which ranged from 2100 grid points for $v=0$ to 90\,000 for
$v=14$. The value of the vibrational wavefunction, provided by LEVEL upon
modification of the input file, was also gathered at each value of $r$. The
expectation value, $\left \langle \psi_v |\Delta V_{\Lambda}(r)|\psi_v \right
\rangle$, was then calculated for each $r$ value and these were summed to
give the overall expectation value for that vibrational level. This was repeated
for all vibrational levels, and these values can be seen in column 2 of
Table~\ref{tab:l-doubling}. Incorporating the $J$ dependence by
evaluating the entirety of Eq.~(\ref{eq:l-doubling}) (multiplying
these corrections by $J(J+1$)) gave the correction for each rotational level
within each vibrational level. These corrections were then simply  added to
the $f$-parity energy to obtain the perturbed $e$-parity energy.

\begin{figure}
    \centering
    \includegraphics[width=0.6\textwidth]{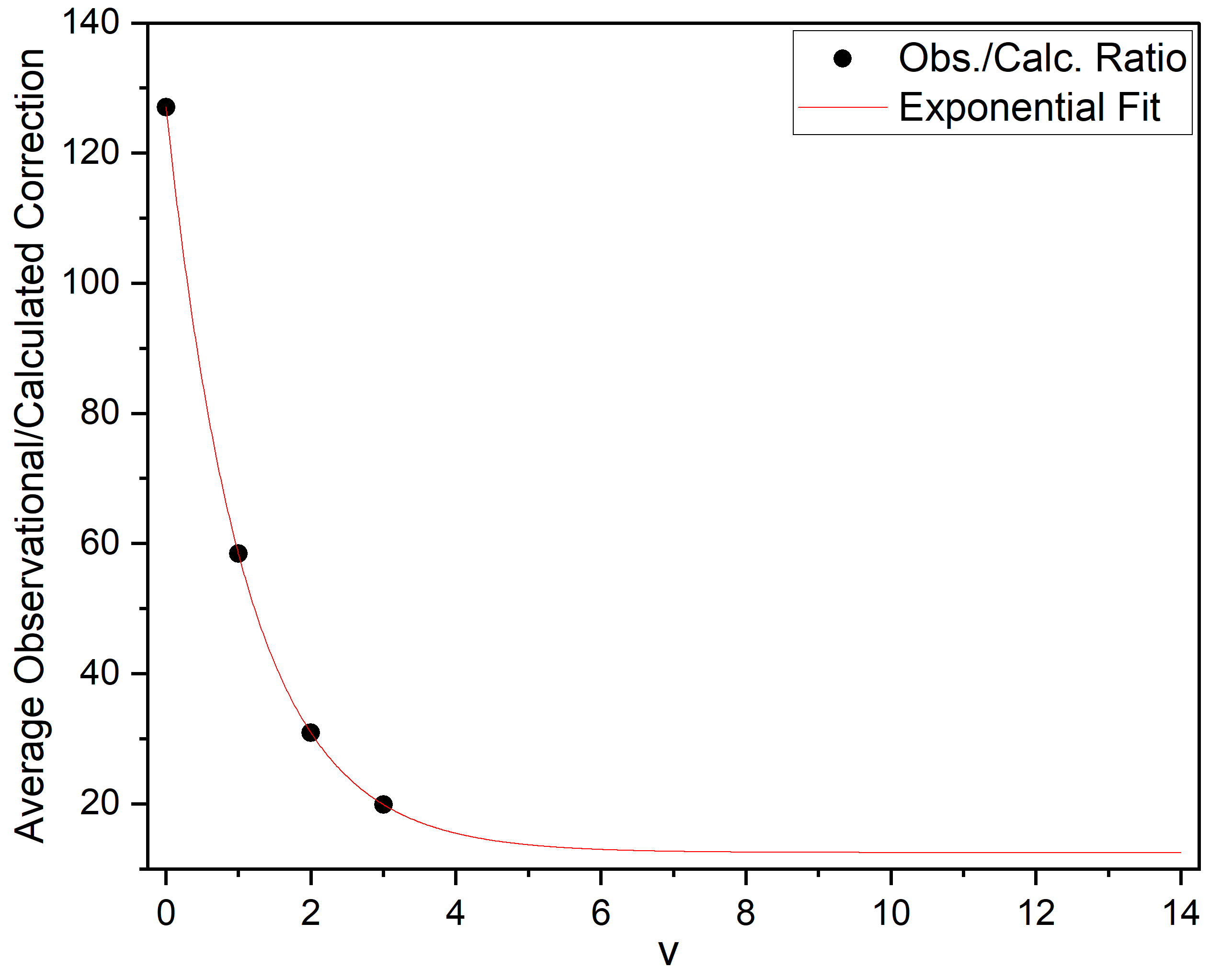}
    \caption{The average ratio of observed to calculated $\Lambda$-doubling
correction ($e-f$ energy) plotted against vibrational level. An extrapolation
to higher vibrational levels is performed using an exponential fit. Values of
average ratio can be seen in Table~\ref{tab:l-doubling} for all $v$
levels.}
    \label{fig:obs-calc-ratios}
\end{figure}

To verify these calculations, comparison was made to observed (experimental)
splittings, some of which were available from MARVEL data (where both $e$ and
$f$ energies for the same rotational level were known). The calculations did
not initially produce results in line with observational corrections, however,
further investigation revealed that the ratio between the calculated and
corresponding observed correction was generally constant within each vibrational
level from $v= 0$ to $v=3$, the region over which the MARVEL data was fairly
comprehensive. Therefore the calculated corrections within each vibrational
level could be scaled by the average ratio to provide an accurate account of
$\Lambda$-doubling in line with observation. The average ratio for each $v$
can be seen in Fig.~\ref{fig:obs-calc-ratios}, showing an apparent
exponentially-decaying trend with increasing $v$. Unfortunately, due to the
limited coverage of the MARVEL data, little information about higher $v$ levels
meant an extrapolation of the curve was necessary to estimate the ratio for higher $v$
levels. The extrapolation was performed using an exponential fit and is shown in
Fig.~\ref{fig:obs-calc-ratios}. From this, the extrapolated ratio for
$v \geq 4$ levels could be estimated, as seen by the asterisked values in
column 3 of Table~\ref{tab:l-doubling}.
With this, the calculated corrections for each level were scaled by the
corresponding ratio to give the final correction for each $v$ level, as can be
seen in the final column of Table~\ref{tab:l-doubling}. After applying
$J$-dependence, the final correction values for each level was
added to the corresponding ($f$-state) energy to give that of the $e$-state.
The scaling with observational corrections means this approach
should be accurate for low $v$'s but the extrapolated corrections
are less secure and would benefit from experimental confirmation.

\begin{table}
\centering
\caption{$\Lambda$-doubling correction calculation parameters, showing the
original expectation value of the $\Lambda$-doubling operator for each
vibrational level, the observed/calculated correction ratio and the resulting
scaled calculated correction. Asterisks indicate extrapolated ratios.
$J$-dependence can then be incorporated by evaluating the entirety of Eq.~(\ref{eq:l-doubling})
using the scaled values to give the final
correction to each $f$-state energy to account for $\Lambda$-doubling.}
\label{tab:l-doubling}
\resizebox{0.5\textwidth}{!}{%
\begin{tabular}{cccc}
\hline
$v$ &\begin{tabular}[t]{@{}c@{}} Original $\left \langle \psi_v |\Delta V_{\Lambda}|\psi_v \right \rangle$ \\ (cm-1) \end{tabular} & \begin{tabular}[t]{@{}c@{}} Average Obs./Calc.\\ Ratio \end{tabular} & \begin{tabular}[t]{@{}c@{}} Scaled $\left \langle \psi_v |\Delta V_{\Lambda}|\psi_v \right \rangle$\\ (cm-1) \end{tabular} \\ \hline
0 & 0.000297 & 127.051 & 0.037787 \\
1 & 0.000594 & 58.449 & 0.034727 \\
2 & 0.001017 & 30.897 & 0.031434 \\
3 & 0.001429 & 19.902 & 0.028440 \\
4 & 0.001736 & 15.464* & 0.026851 \\
5 & 0.001906 & 13.690* & 0.026096 \\
6 & 0.001946 & 12.978* & 0.025261 \\
7 & 0.001877 & 12.693* & 0.023827 \\
8 & 0.001713 & 12.579* & 0.021553 \\
9 & 0.001463 & 12.533* & 0.018334 \\
10 & 0.001134 & 12.515* & 0.014188 \\
11 & 0.000752 & 12.507* & 0.009402 \\
12 & 0.000384 & 12.504* & 0.004800 \\
13 & 0.000136 & 12.503* & 0.001704 \\
14 & 0.000026 & 12.503* & 0.000322 \\ \hline
\end{tabular}%
}
\end{table}

\subsubsection{Einstein Coefficients}

LEVEL  computes Einstein $A$ coefficients using the expression
\begin{equation}
\label{eq:Einstein}
    A_{v',J',v'',J''} = 3.1361891\times 10^{-7}\:\frac{S(J',J'')}{2J'+1}\:\nu^{3}\:|\left
\langle \Psi_{v',J'}|M(r)|\Psi_{v'',J''} \right \rangle|^2
\end{equation}
where $A$ is the Einstein coefficient (s$^{-1}$), $S(J',J'')$ is the
H\"{o}nl-London rotational intensity factor \citep{HANSSON2005169}, $J'$ the
upper level rotational quantum number, $\nu$ the transition wavenumber
(cm$^{-1}$), $\Psi_{v',J'}$ and $\Psi_{v'',J''}$ the normalised initial
and final state radial wavefunctions and $\left \langle
\Psi_{v',J'}|M(r)|\Psi_{v'',J''} \right \rangle$ the expectation value of the
dipole/transition dipole moment function $M(r)$, which we asked LEVEL to output.

LEVEL's lack of consideration of $\Lambda$-doubling means $e$-parity A\,$^{1}\Pi$ levels are not present and
transitions involving these states were not correctly accounted for.
A\,$^{1}\Pi$($e$)-X transitions were incorrectly assigned the wavenumber of
the corresponding (forbidden by dipole selection rules) A\,$^{1}\Pi$($f$)-X
transition which led to an incorrect calculation of the Einstein $A$ coefficient
since  Eq.~(\ref{eq:Einstein}) depends on the transition wavenumber.
By applying the relevant
$e-f$ correction to each  transition wavenumber, the Einstein coefficient was recalculated using
Eq.~(\ref{eq:Einstein}) and the outputted dipole matrix elements.
For transitions within the A\,$^{1}\Pi$ state this required some care due to many transitions being unaccounted for, as there exist potentially four transitions where LEVEL computes just one transition and thus one $A$ coefficient that uses (often incorrect) transition information that LEVEL provided.
For example, LEVEL gives Q-branch ($\Delta J$=0) transitions within the same
vibrational level a wavenumber of 0 cm$^{-1}$, but in reality these correspond
to long-wavelength, allowed $\Lambda$-doublet transitions between $e$ and $f$ states of
the same rotational level. This meant calculating
Eq.~(\ref{eq:Einstein}) in its entirety including determining
the relevant H\"{o}nl-London factor \citep{HANSSON2005169}. Note that these
methods to recalculate the Einstein coefficient assume the expectation value of
the DM/TDM function in Eq.~(\ref{eq:Einstein}) remains unchanged, which is in
any case a common approximation used in other treatments and is unlikely to be problematic.

\subsection{Final line lists}

This above calculation procedure was followed for both $^{12}$CH$^+$ and $^{13}$CH$^+$ and
line lists were generated considering states with $J$ up to 68 and 69, respectively.
For $^{12}$CH$^+$ one further step was performed. The MARVEL energy levels with
associated uncertainties were used to replace the ones computed by LEVEL.

\begin{figure}
    \centering
        \includegraphics[width=.51\textwidth]{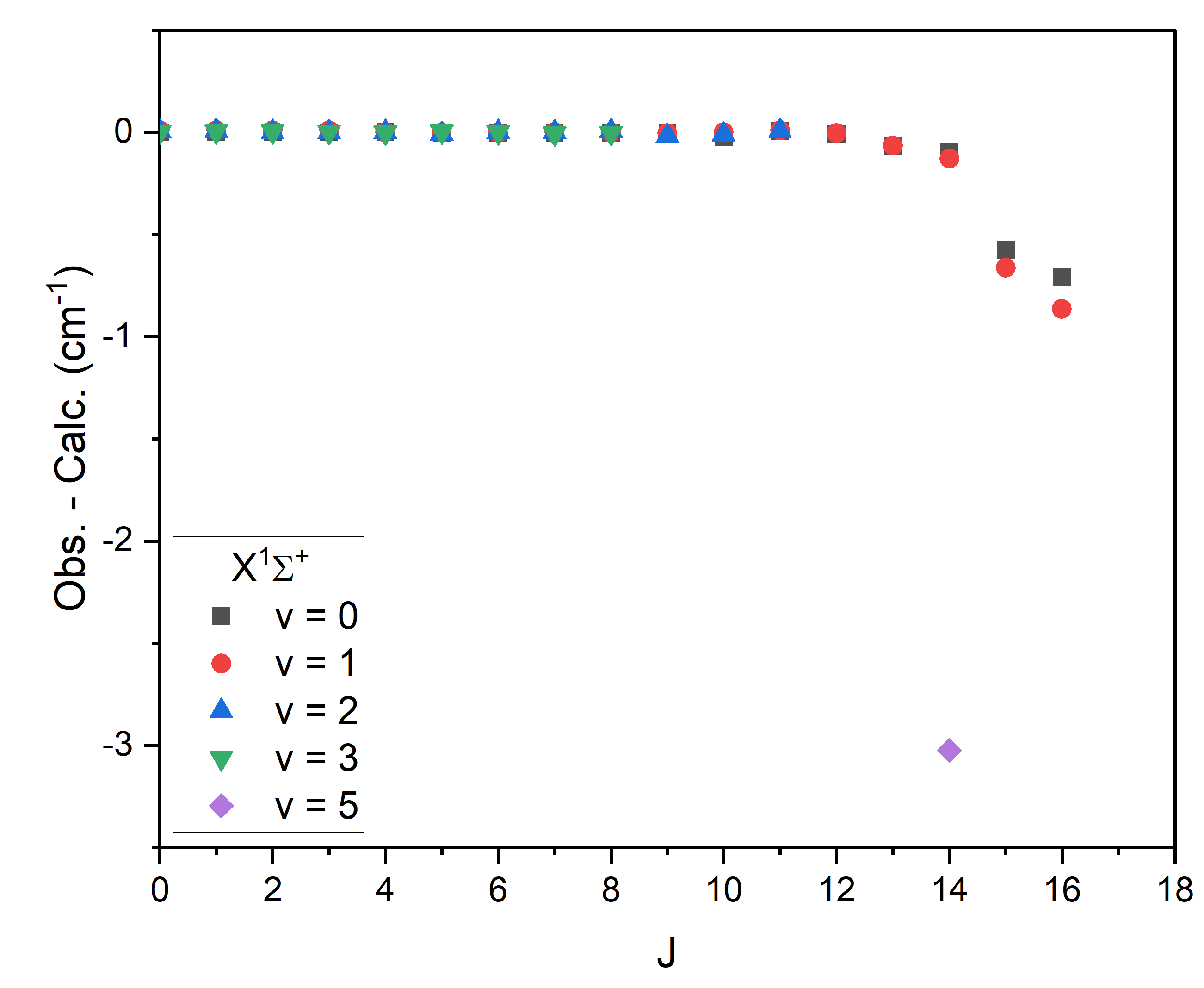}
        \includegraphics[width=.5\textwidth]{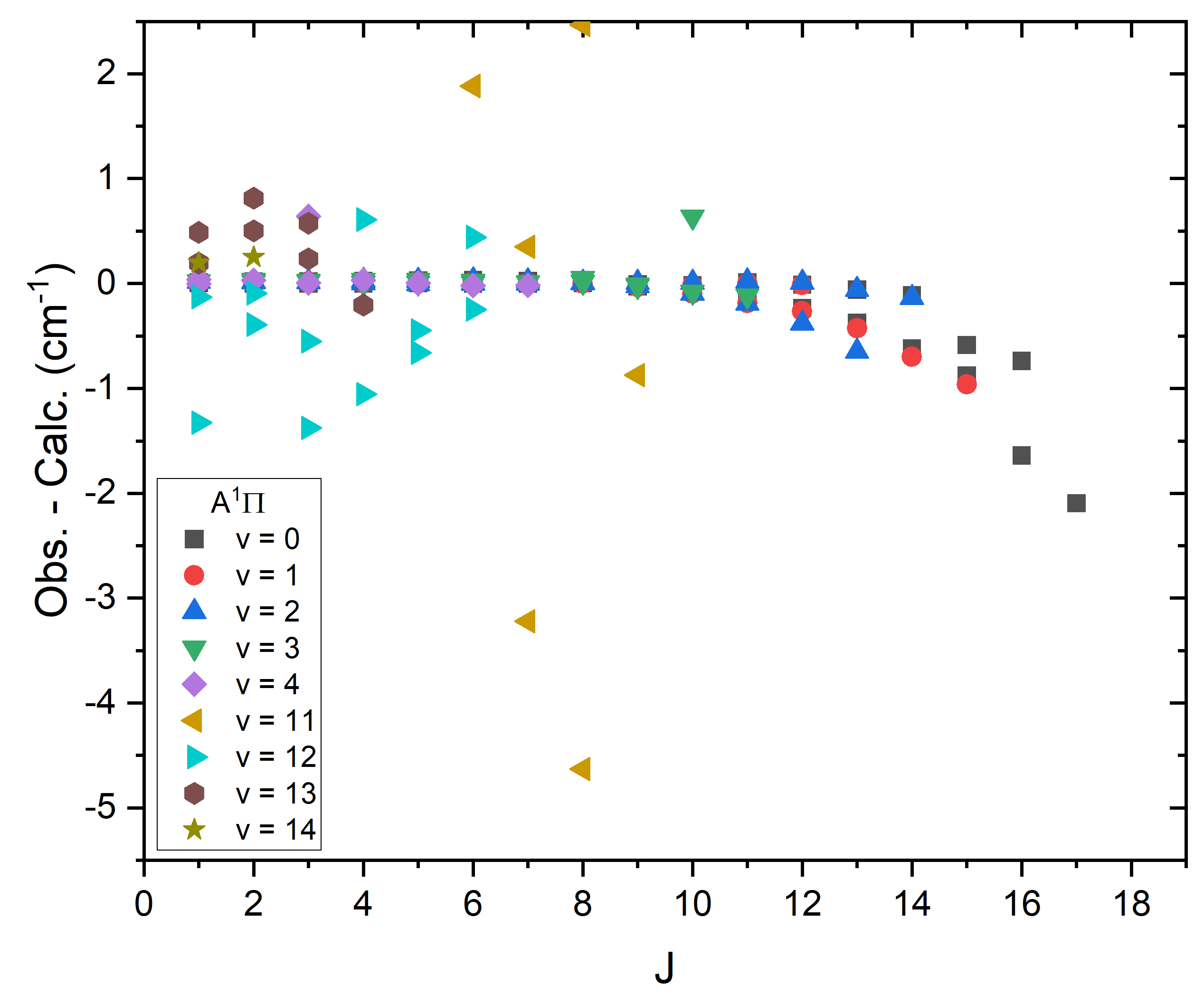}
    \caption{Obs. -- Calc. residuals for the X\,$^{1}\Sigma^{+}$ (Top) and
A\,$^{1}\Pi$ (Bottom) states for different vibrational levels as a function of
$J$. Note that the two values for each ($v, J$) level in the A\,$^{1}\Pi$
plot are the $e$ and $f$ parity states, split by $\Lambda$-doubling.}
    \label{fig:obs-calc}
\end{figure}

Due to the greater accuracy of the empirical MARVEL data, calculated $^{12}$CH$^+$  energies
were replaced with the respective MARVEL values where available in the States
file. These `MARVELised' energy levels can be identified via the `Label' column in the states file
which indicates `Ma' if the MARVEL energy level was used, or `Ca' if there was
no such available data. Note that the calculated energy prior to MARVELisation
is given in the `Calc.' column. The uncertainty in the States file is determined
depending on the energy level used. For the MARVELised energies, the uncertainty
provided by MARVEL was simply used. For the calculated energies (which includes all $^{13}$CH$^+$ levels),
an estimate for the uncertainty was made using the
following expression:
\begin{equation}\label{eq:uncertainty}
    \textrm{uncertainty} = av + bJ(J+1)
\end{equation}
where $a=0.05/0.1$ cm$^{-1}$ and $b=0.01/0.01$ cm$^{-1}$ for are constant parameters for the
X\,$^{1}\Sigma^{+}$/A\,$^{1}\Pi$ electronic
states, respectively. These parameters were estimated by analysing the evolution of $^{12}$CH$^+$ obs. $-$
calc. energy level residuals seen in Fig.~\ref{fig:obs-calc}, which
plots the agreement of the MARVEL (obs.) and LEVEL (calc.) data and how
this varies with $v$ and $J$ within each electronic state. Overall, the
residuals were very small for low $v$ and/or
$J$ levels of both the electronic states. Naturally, A\,$^{1}\Pi$ has
greater residuals, especially for high $v$, since higher energies are probed
and as a result of the manual treatment of $\Lambda$-doubling, but in general
both states exhibit good accuracy. An apparent quadratic $J$ dependence can be
seen in the residuals in Fig.~\ref{fig:obs-calc}, motivating the form
of Eq.~(\ref{eq:uncertainty}).


\section{Results}
\subsection{Line list and File Structure}
With calculated energies for rotational and vibrational levels of
X\,$^{1}\Sigma^{+}$ and A\,$^{1}\Pi$, and transition frequencies and
Einstein coefficients for allowed transitions, the components of the line list
of CH$^{+}$ were complete. In accordance with ExoMol format, this was
reformatted into a States and Transitions file \citep{jt548}, extracts from
which for $^{12}$CH$^+$ can be seen in Tables~\ref{tab:states} and \ref{tab:trans}
respectively. The $^{12}$CH$^+$ States file contains 1505
rovibrational energy levels covering the X\,$^{1}\Sigma^{+}$ and
A\,$^{1}\Pi$ electronic states up to a few thousand wavenumbers above the
dissociation limit, additionally providing their uncertainty and associated
quantum numbers defining the level. The Transitions file, characterising 34\,194
transitions, contains Einstein $A$ coefficients and frequencies for each allowed transition between the energy levels of the States file. For $^{13}$CH$^+$ there are 1519 states and 42\,387 transitions.
We call these line lists PYT.

\begin{table}
{\tt
\centering
\caption{Extract from the States file for $^{12}$CH$^+$.}
\label{tab:states}
\resizebox{\textwidth}{!}{%
\begin{tabular}{ccrrrrcclrrrrcr}
\hline
$i$ & \multicolumn{1}{c}{{\rm Energy} ({\rm cm}$^{-1}$)} & $g_{i}$ & $J$ & \multicolumn{1}{c}{{\rm Unc.} ({\rm cm}$^{-1}$)} & $\tau$ (s) & $+/-$ & $e/f$ & \multicolumn{1}{c}{{\rm State}} & $v$ & $\Lambda$ & $\Sigma$ & $\Omega$ & \multicolumn{1}{c}{{\rm label}}& \multicolumn{1}{c}{{\rm Calc.}} \\
\hline
    367&    24005.218550&    134&     33&       11.470000&  1.1470E+01 &  $-$ &   e  & X1Sigma+     &    5&    0&    0&    0& Ca  &    24005.218550\\
    368&    24038.829766&     10&      2&        0.610000&  6.1000E-01 &   +  &   e  & X1Sigma+     &   11&    0&    0&    0& Ca  &    24038.829766\\
    369&    24072.559047&     26&      6&        0.334901&  3.3490E-01 &  $-$ &   f  & A1Pi         &    0&    1&    0&    1& Ma  &    24072.560989\\
    370&    24074.180657&     26&      6&        0.217601&  2.1760E-01 &   +  &   e  & A1Pi         &    0&    1&    0&    1& Ma  &    24074.148026\\
    371&    24091.232882&     14&      3&        0.670000&  6.7000E-01 &  $-$ &   e  & X1Sigma+     &   11&    0&    0&    0& Ca  &    24091.232882\\
    372&    24107.338107&     78&     19&        4.250000&  4.2500E+00 &  $-$ &   e  & X1Sigma+     &    9&    0&    0&    0& Ca  &    24107.338107\\
    373&    24131.807092&    110&     27&        7.910000&  7.9100E+00 &  $-$ &   e  & X1Sigma+     &    7&    0&    0&    0& Ca  &    24131.807092\\
    374&    24160.970652&     18&      4&        0.750000&  7.5000E-01 &   +  &   e  & X1Sigma+     &   11&    0&    0&    0& Ca  &    24160.970652\\
    375&    24162.245191&    146&     36&       13.520000&  1.3520E+01 &   +  &   e  & X1Sigma+     &    4&    0&    0&    0& Ca  &    24162.245191\\
    376&    24165.617170&    166&     41&       17.320000&  1.7320E+01 &  $-$ &   e  & X1Sigma+     &    2&    0&    0&    0& Ca  &    24165.617170\\
    377&    24229.503713&     58&     14&        2.600000&  2.6000E+00 &   +  &   e  & X1Sigma+     &   10&    0&    0&    0& Ca  &    24229.503713\\
    378&    24229.561434&     30&      7&        0.366501&  3.6650E-01 &   +  &   f  & A1Pi         &    0&    1&    0&    1& Ma  &    24229.563130\\
    379&    24231.703209&     30&      7&        0.204901&  2.0490E-01 &  $-$ &   e  & A1Pi         &    0&    1&    0&    1& Ma  &    24231.679179\\
    380&    24247.929131&     22&      5&        0.850000&  8.5000E-01 &  $-$ &   e  & X1Sigma+     &   11&    0&    0&    0& Ca  &    24247.929131\\
    381&    24351.966003&     26&      6&        0.970000&  9.7000E-01 &   +  &   e  & X1Sigma+     &   11&    0&    0&    0& Ca  &    24351.966003\\
    382&    24408.044035&     34&      8&        0.354901&  3.5490E-01 &  $-$ &   f  & A1Pi         &    0&    1&    0&    1& Ma  &    24408.042265\\
 \hline
\end{tabular}%
}
}
$i$: State counting number, $g_{i}$: Total statistical weight, $J$: Total angular momentum, Unc: uncertainty (cm$^{-1}$), $\tau$: lifetime (s), $+/-$: Total parity, $e/f$: Rotational parity, $v$: Vibrational level quantum number, $\Lambda$: Projection of electronic angular momentum, $\Sigma$: Projection of electron spin,  $\Omega$: Projection of total angular momentum ($\Omega = \Lambda + \Sigma$), Label: `Ma' denotes MARVEL energy level, `Ca' denotes LEVEL calculated energy level, Calc.: LEVEL calculated energy level value.

\end{table}

\begin{table}
\centering
{\tt
\caption{Extract from the Transitions file for $^{12}$CH$^{+}$.}
\label{tab:trans}
\resizebox{0.4\textwidth}{!}{%
\begin{tabular}{rrrl}
\hline
$f$    & $i$   & $A_{fi}$ (s$^{-1}$) & $\tilde{\nu}_{fi}$ ({\rm cm}$^{-1})$       \\ \hline
516 & 377 & 3.1861E+01 & 2754.475947 \\
1178 & 759 & 1.1669E-09 & 2755.165546 \\
92 & 58 & 1.5055E+00 & 2756.287971 \\
658 & 473 & 3.9484E+00 & 2756.308088 \\
544 & 392 & 2.8961E+01 & 2756.981308 \\
266 & 210 & 2.5046E+01 & 2757.306730 \\
1169 & 754 & 5.0213E-09 & 2757.533295 \\
1161 & 748 & 1.9419E-02 & 2757.544843 \\
991 & 674 & 2.2755E-03 & 2757.621535 \\
612 & 440 & 1.0042E+01 & 2758.338556 \\
1166 & 751 & 1.6578E-09 & 2758.406401 \\
186 & 141 & 1.7163E+01 & 2758.513693 \\
1298 & 774 & 1.1399E+01 & 2758.878845\\ \hline
\multicolumn{4}{l}{\begin{tabular}[c]{@{}l@{}}$f$: Final state counting number,\\ $i$: Initial state counting number,\\ $A_{fi}$: Einstein $A$ coefficient (s$^{-1}$),\\ $\tilde{\nu}_{fi}$: Transition wavenumber (cm$^{-1}$). \end{tabular}}
\end{tabular}%
}
}
\end{table}

It should be noted that, in the $^{12}$CH$^+$ Transitions file, all A-A transitions involving vibrational levels with
$v \geq 4$ levels are excluded. This is due to further issues with LEVEL,
which gave an error for these transitions.  No A-A transitions have ever been
detected and they are unlikely to be seen astronomically in the foreseeable future so this
is not a significant defect.

\subsection{Partition Functions}
\begin{figure}
    \centering
    \includegraphics[width=0.5\textwidth]{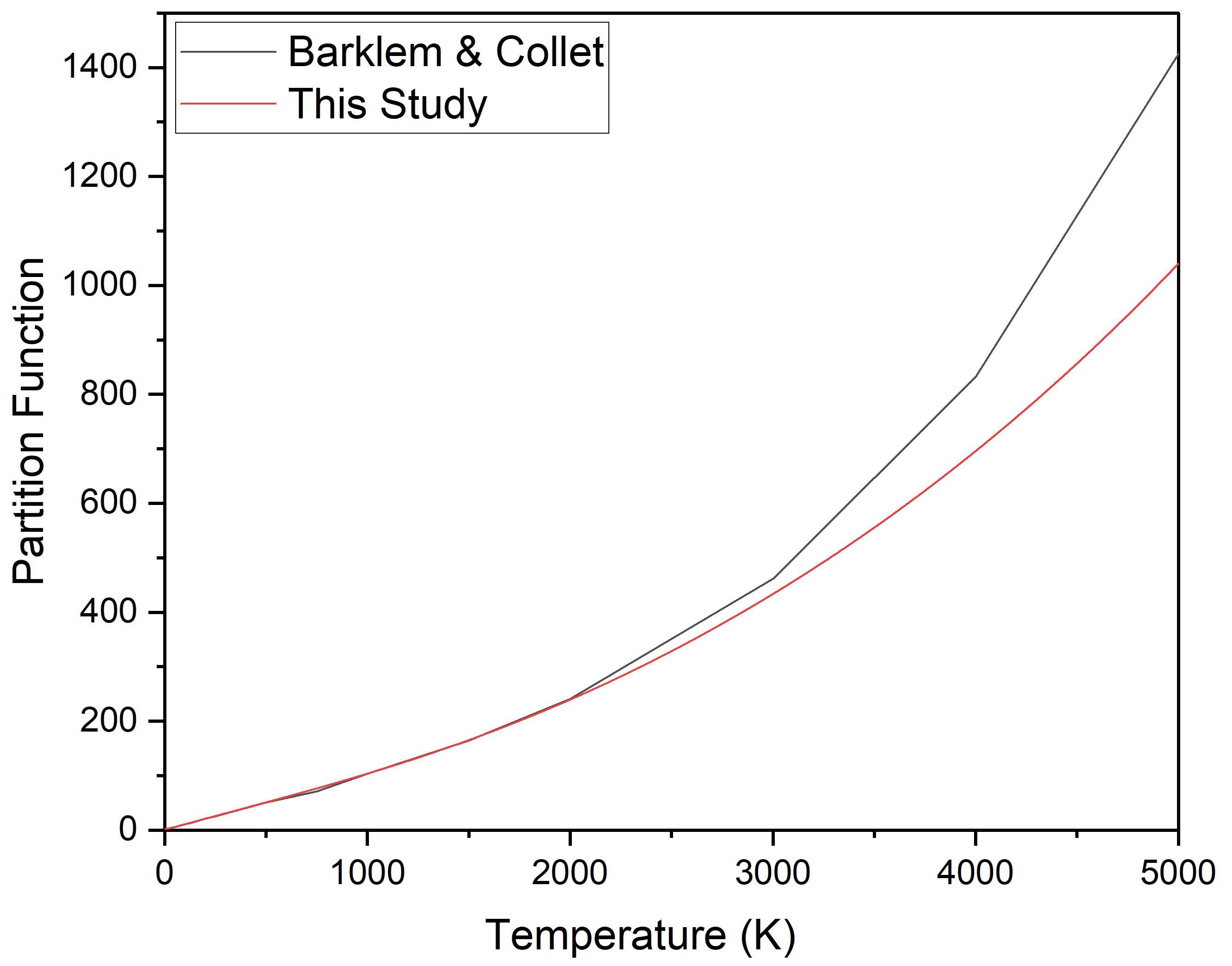}
    \caption{CH$^{+}$ partition function as calculated from the line list by
ExoCross compared with that of  \citet{16BaCoxx.partfunc}.}
    \label{fig:partfunc}
\end{figure}

A partition function for $^{12}$CH$^{+}$ was provided by \citet{16BaCoxx.partfunc} for 42
temperatures up to 10\,000 K. The program ExoCross \citep{jt708} was used to calculate the partition
function for both $^{12}$CH$^{+}$ and $^{13}$CH$^{+}$ in steps of 1 K up to 10\,000 K. The
ExoMol convention, in accordance with HITRAN \citep{jt692}, is to provide
partition functions that include full atomic nuclear spin degeneracy,
$g_{ns}$. For $^{12}$C$^{1}$H$^{+}$, this gives a value of 2, while for  $^{13}$CH$^{+}$ it gives 4. As such,
the values of the $^{12}$CH$^{+}$ partition function of Barklem and Collet, who do not account
for this, have been multiplied by 2 for comparison, which can be seen in
Fig.~\ref{fig:partfunc}. Very good agreement can be seen below about
3000 K. At higher temperatures, agreement decreases somewhat which is likely
due to the increasingly significant thermal occupation of higher energy
electronic states that were not considered in this study, most notably the
a\,$^{3}\Pi$ state which lies between X\,$^{1}\Sigma^{+}$ and
A\,$^{1}\Pi$ 
As a result, a maximum temperature of 5000 K has been shown. JPL
also provided a $^{12}$CH$^{+}$ partition function for several low temperatures, the largest of
which, at 300 K, 30.7006 \citep{JPL} (again after accounting for nuclear spin
degeneracy) shows excellent agreement with the calculated value here of 30.6990.

\subsection{Spectra}
\begin{figure}
    \centering
        \includegraphics[width=.5\textwidth]{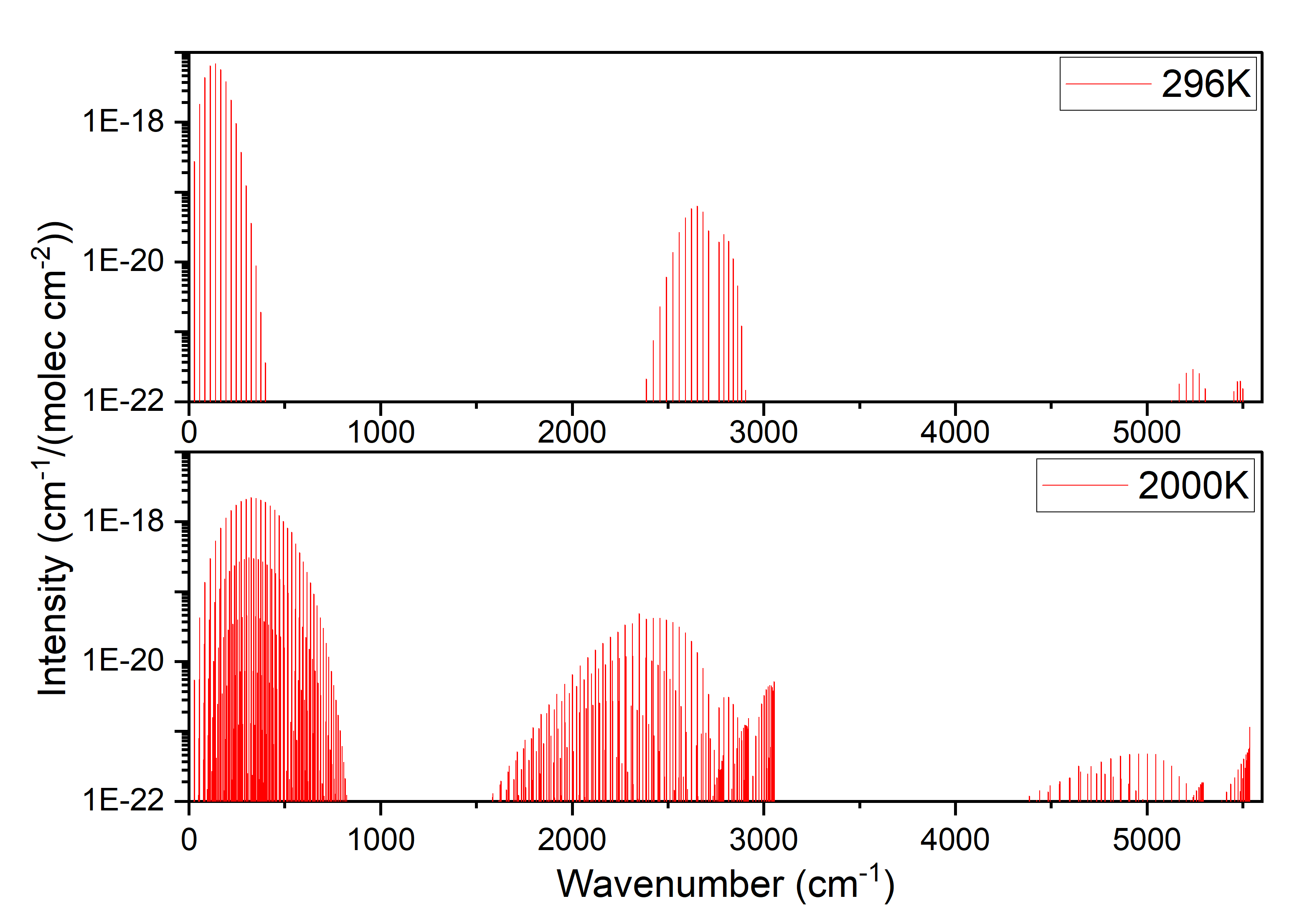}
        \includegraphics[width=.5\textwidth]{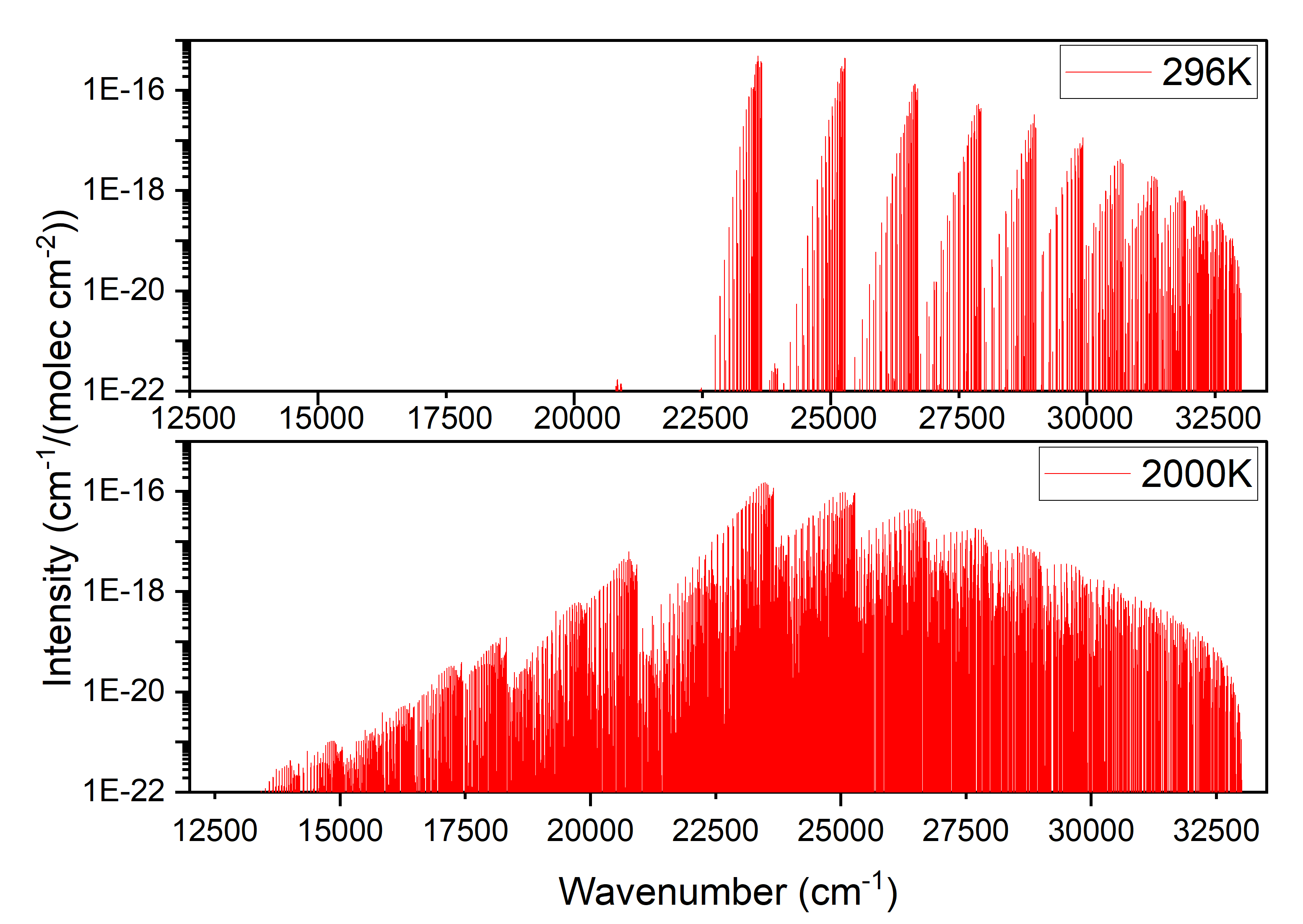}
    \caption{Simulated absorption spectra at 296 K and 2000 K, covering the rotational and vibrational (IR) bands (Top) and the electronic bands (Bottom). Spectra produced using a Doppler broadened profile with ExoCross.}
    \label{fig:overviewspectra}
\end{figure}

Using the PYT $^{12}$CH$^+$ line list, ExoCross \citep{jt708} was used to simulate
absorption/emission spectra at various temperatures.
Figure~\ref{fig:overviewspectra} shows simulated absorption spectra for
temperatures of 296 K and 2000 K. A Doppler broadened Gaussian profile that is
dependent on temperature has been used. The spectra have been separated with
rotational and vibrational spectra in the top panel and electronic in the
bottom, to avoid large gaps in the plot. Note that a logarithmic intensity scale
has been used to show the relatively weak vibrational (IR) spectrum (seen for the
wavenumber $> 1000$ cm$^{-1}$ region of the top-panel plots) compared to the
rotational and electronic spectrum. This corroborates the elusiveness of this
spectrum as noted by previous studies.

To assess the quality of the calculated line list, comparison with previously recorded spectra are made.
As no prior line list for CH$^{+}$ exists, nor an
experimental/observational spectrum covering an extensive wavelength range,
comparison are made to selected  portions of the
full spectrum.

\begin{figure}
    \centering
    \includegraphics[width=0.5\textwidth]{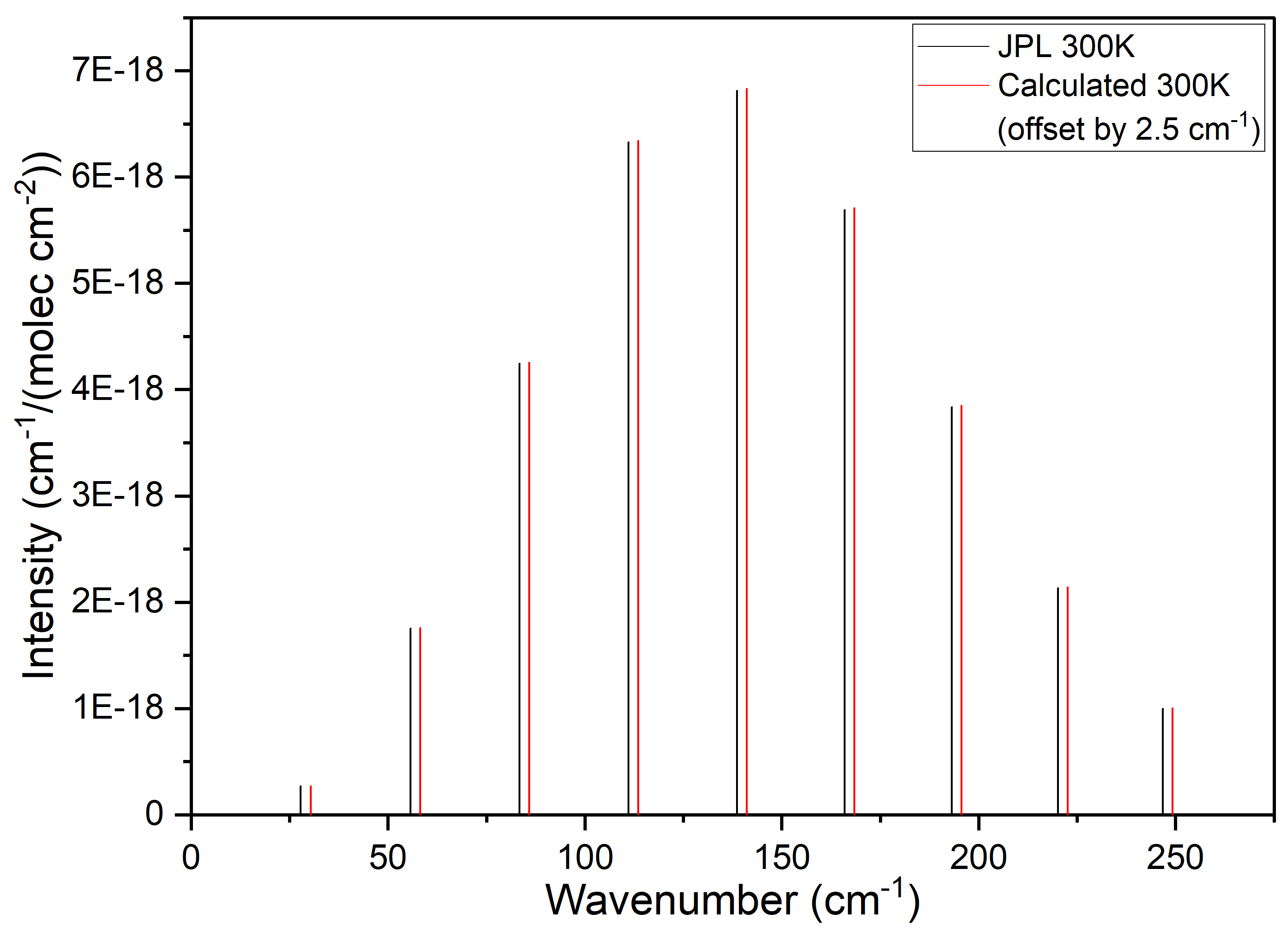}
    \caption{Comparison of the JPL CH$^{+}$ X\,$^{1}\Sigma^{+}$ rotational spectrum (black) with the calculated rotational stick spectrum from this study (red). Note that the calculated data has been offset by 2.5 cm$^{-1}$ for intensity comparison.}
    \label{fig:RotComp}
\end{figure}

Low-lying pure rotational transitions of CH$^{+}$ have been previously
measured, as seen in Table~\ref{tab:observation}, but no rotational
spectra are presented. However, the JPL spectroscopy database \citep{JPL}, who
also aim to provide spectroscopic data for molecular transitions, used the first
laboratory detection of the $J=1-0$ CH$^{+}$ rotational line \citep{06PeDr}
along with previous electronic A-X band data to generate a predicted rotational
spectrum up to $J=8$. Although, as discussed previously, this laboratory
detection was later found to be inaccurate by 0.0019 cm$^{-1}$, and a wider
range of rotational lines have since been detected both in the lab and in space,
JPL has not updated their database and the inaccurate data is still presented.
Nevertheless, this discrepancy would not be visible on the scale of comparing
multiple lines and their intensities should still be accurate, as such the JPL
rotational spectrum of CH$^{+}$ has been compared with the calculations of
this study in Fig.~\ref{fig:RotComp}. The simulated stick spectrum
from ExoCross was run at a temperature of 300 K to match that of JPL, and the
JPL intensities were converted to standard ExoMol/HITRAN units of cm/molecule for comparison. In Fig.~\ref{fig:RotComp},
the calculated spectrum of this study has been offset by +2.5
cm$^{-1}$ to allow for visual comparison, as the wavenumbers agree to a high
degree of accuracy. The intensities can also be seen to agree
closely.

\begin{figure}
    \centering
    \includegraphics[width=0.75\textwidth]{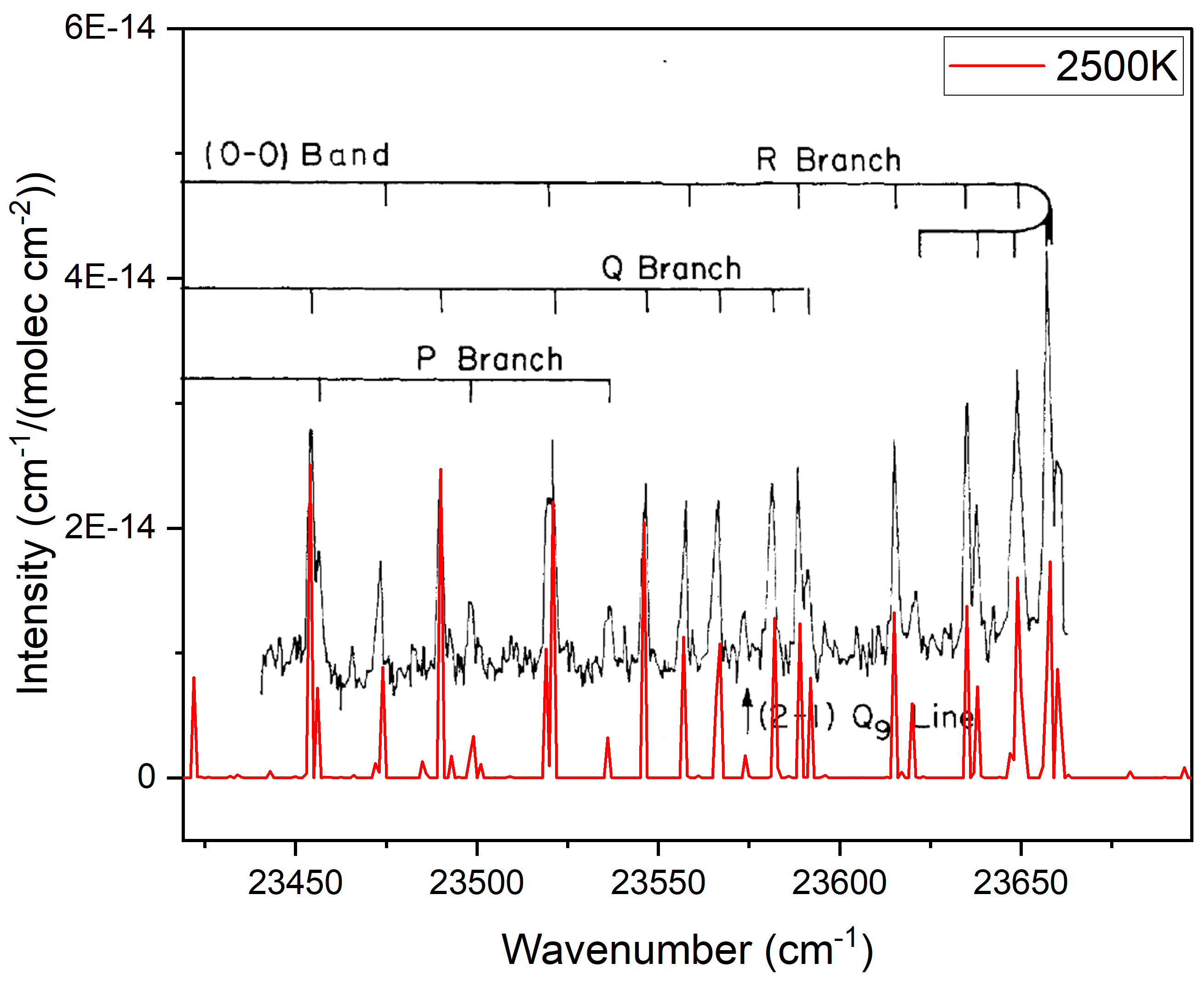}
    \caption{Comparison of the \protect{$v=0-0$ A\,$^{1}\Pi$ -- X\,$^{1}\Sigma^{+}$}
band spectrum from  \citet{81GrBrOkWi} (black) with the simulated
emission spectrum from ExoCross at 2500K (red). A Doppler line profile was used.
The intensity scale corresponds only to the calculated ExoCross data, since
Grieman et al. provided arbitrary units.}
    \label{fig:81ElecComp}
\end{figure}

The weak nature of the vibrational (IR) spectrum of CH$^{+}$ means it has only
been detected in one astrophysical object \citep{21NeGoChFa} and one laboratory
study \citep{18DoJuScAs}, of which no presented spectra exist, so visual
comparison cannot be made. The A-X electronic spectrum is by far the most
studied of CH$^{+}$ with many experimental studies focusing on the molecule,
however, presented spectra are limited. A study by
\citet{81GrBrOkWi} used laser-induced fluorescence to probe CH$^{+}$ emission
and presented a portion of the $v=0-0$ band of the A-X spectrum, a comparison
to which, using a simulated emission spectrum, has been shown in
Fig.~\ref{fig:81ElecComp}. The previous study also gave arbitrary intensity units and is probably not recorded in
sample in thermodynamic equilibrium, so
the intensity pattern is not expected to match that of this study. As such,
a temperature of 2500 K was somewhat arbitrarily chosen for our calculated
spectrum based on the high temperature requirement and certain similarities in
intensity pattern. On Fig.~\ref{fig:81ElecComp}, the intensity units
correspond only to the results of this study (red). The spectrum from the
previous study had to be offset slightly to match the spectrum of this study
which indicates a slight error in their axes, since the transition wavenumbers
of the two studies match to a high degree of accuracy.

\begin{figure}
    \centering
    \includegraphics[width=0.75\textwidth]{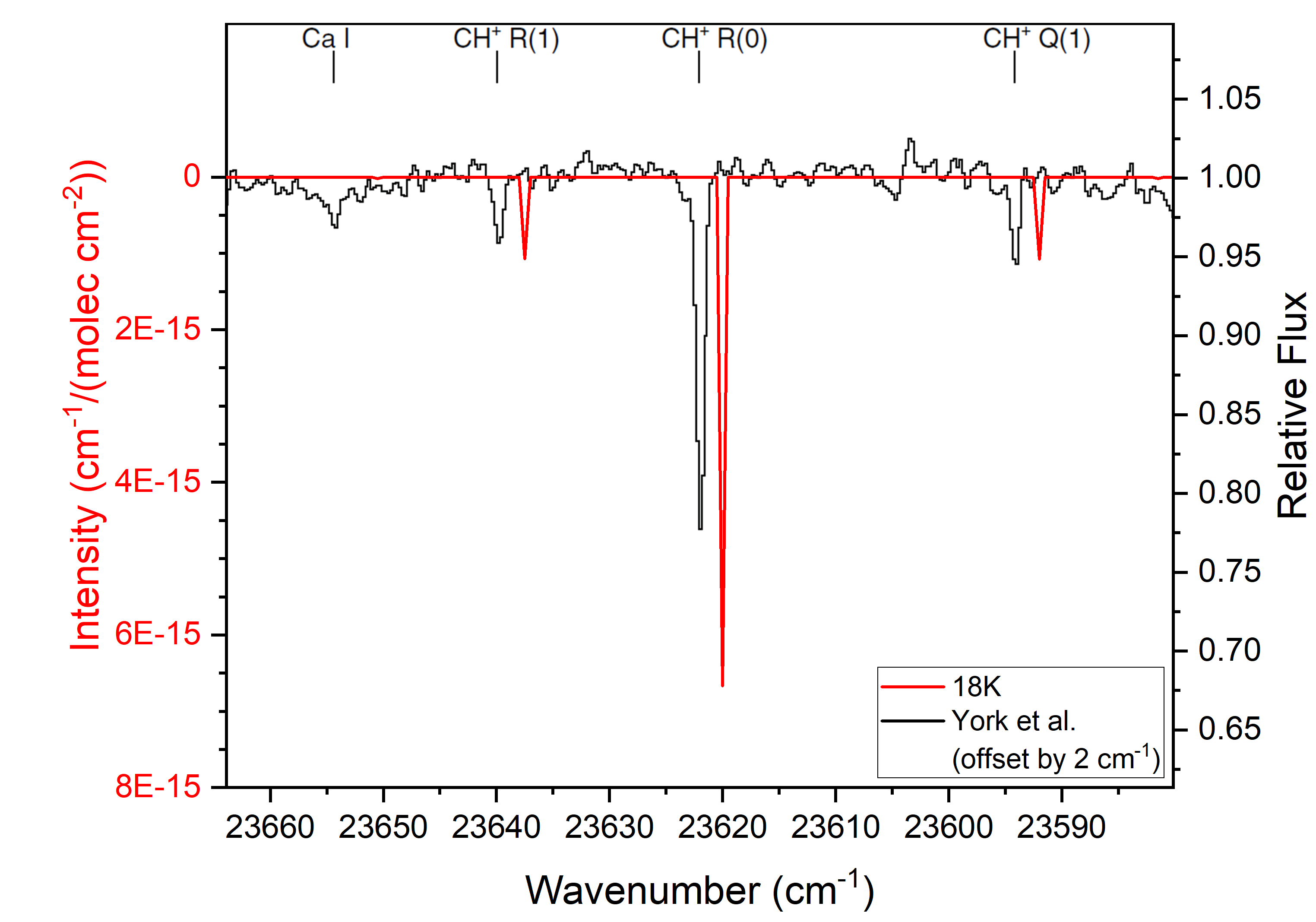}
    \caption{Comparison of the A-X spectral lines of the diffuse interstellar
bands (DIBs) observed towards Herschel 36 (black) \citep{13YoDaWe.CH+} with the
simulated absorption spectrum of this study at 18 K (red), for which a Doppler
line profile is used. The red units on the left axis correspond to the
calculated data of this study, while the black units on the right correspond to
the intensity units provided by the DIB study. Note that wavenumber increases
right to left as the original DIB study used wavelength units. The DIB data was
originally blueshifted by \protect{$\sim +6.75$ cm$^{-1}$ ($-$1.21 \AA), after
aligning the two spectra the DIB data was offset by +2 cm$^{-1}$} (to the
left) for clearer intensity comparison. }
    \label{fig:14ElecComp}
\end{figure}

CH$^{+}$ A-X absorption was seen in diffuse interstellar bands (DIBs) along
the line of sight towards star system Herschel 36 and three lines were detected
\citep{13YoDaWe.CH+}, R(0), R(1) and Q(1). A comparison to the spectra of these
is shown in Fig.~\ref{fig:14ElecComp}. Note that since the study used
wavelength units for their spectrum, wavenumber increases right to left. Upon
analysis, the DIBs were observed to be blueshifted relative to the calculated
lines with an offset of approximately +6.75 cm$^{-1}$
($-1.21$ \AA, indicating the absorbing object is moving at 85.7 km s$^{-1}$
towards us). After matching the line positions, an offset of +2 cm$^{-1}$ (to
the left) was applied to the DIB data to allow for visual comparison of the
intensities. The red intensity units correspond to calculated data, while the
black relative flux units correspond to the DIB study.
Since the temperature of the absorption region was not found in the previous
study, spectra were simulated at various temperatures using ExoCross to
match up the intensity pattern of the observed lines, which would be expected
to agree between these two studies. It appears that a temperature of 18 K
matches the $J=1$ lines well (by eye) while overestimating the observed
$J=0$ line flux. A temperature
higher than 18 K results in an intensity decrease from $J=0$ levels but an increase
from $J=1$ levels (it is assumed that thermal population of a rotational level
is proportional to the intensity of transitions from that level). Thus the
intensity of the $J=0$ and $J=1$ lines on the above spectra cannot both
match up. This suggests that the region is consistent with a temperature of 18
K, but that the observed $J=0$ line is optically thick, and thus is limited in
its intensity.

\subsection{Einstein Coefficients}
\begin{figure}
    \centering
    \includegraphics[width=0.6\textwidth]{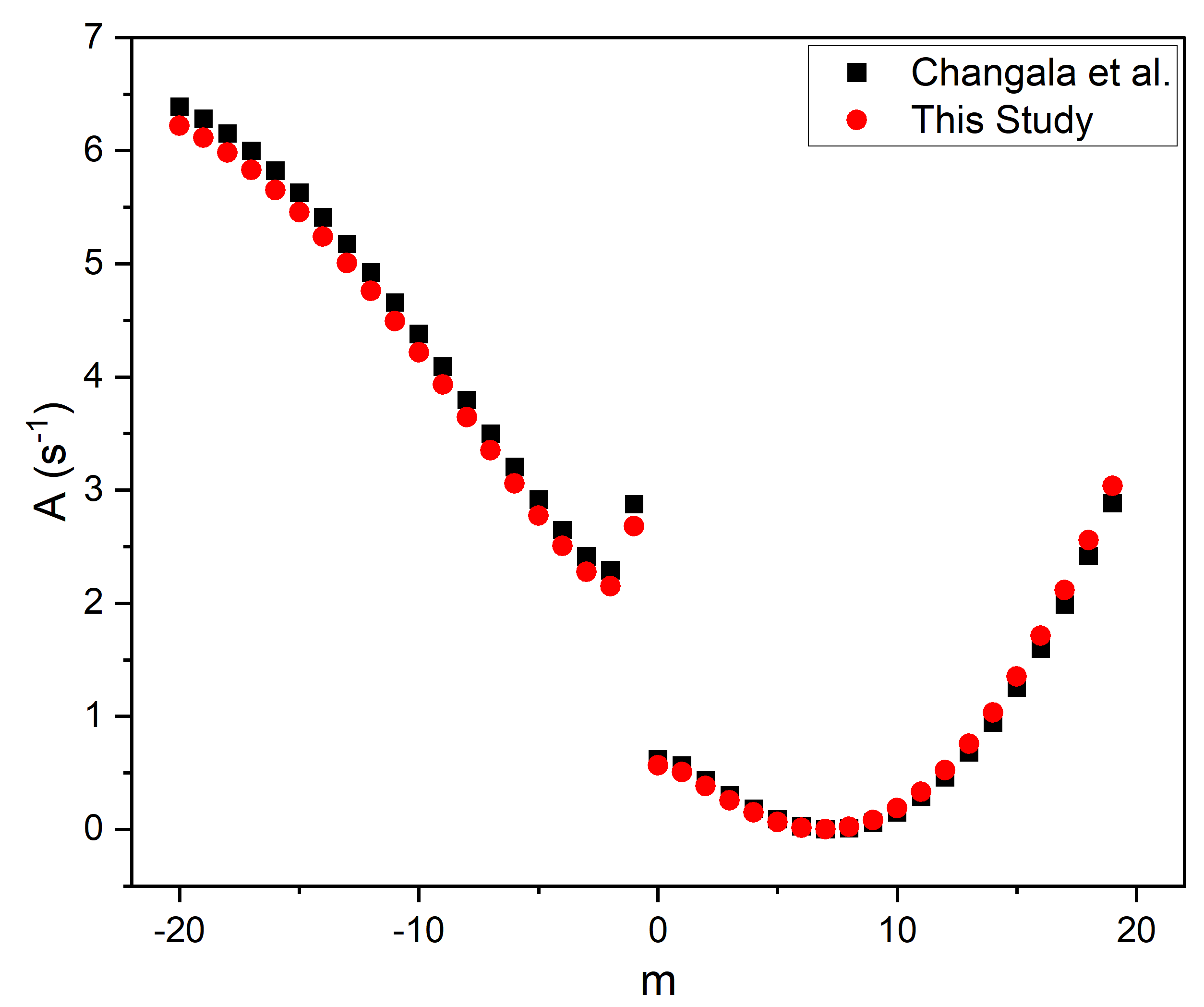}
    \caption{Comparison of Einstein coefficients for the \protect{$v=1-0$ band of
X\,$^{1}\Sigma^{+}$ from  \citet{21ChNeGo.CH+} with those
calculated in this study. P and R-branch transitions have been organised into a
parameter $m$ for plotting, where the value of $m$ indicates the (lower
state) rotational quantum number of the transition, and the $\pm$} indicates
R/P-branch respectively}
    \label{fig:Einstein}
\end{figure}

Einstein $A$ coefficients can be compared with the work of  \citet{21ChNeGo.CH+} who analysed unusual
CH$^{+}$ rovibrational emission patterns in NGC 7027 and gave Einstein
coefficients for R- and P- branch transitions in the $v=1-0$ X-X band in an
attempt to explain unexpectedly weak R-branch transitions. The calculated
coefficients of this study have been compared with these in Table~\ref{tab:Einstein} and Figure~\ref{fig:Einstein}. For
plotting, a parameter $m$ has been used, positive $m$ values indicate
R-branch transitions, and negative ones indicate P-branch transitions,
originating from the rotational state $J=m$. The Einstein coefficients are
seen to agree well, which supports the observation of weaker R-branch
transitions.  \citet{21ChNeGo.CH+} also used PECs from
\citet{16ChLe.CH+}, so the differences in results likely lies
with differences in fitting, or the differing dipole moments used. Despite
this, since the DMs for these vibrational transitions are so small (hence the
weakness of the observed transitions), the level of disagreement between results
is unproblematic.

\begin{table}
\centering
\caption{Comparison of Einstein A coefficients for the \protect{$v=1-0$ band of
X\,$^{1}\Sigma^{+}$} from  \citet{21ChNeGo.CH+} to those of this  study.}
\label{tab:Einstein}
\resizebox{\textwidth}{!}{%
\begin{tabular}{lllllllll}
\hline
\textbf{Transition} & \textbf{Changala et al.} & \multicolumn{1}{l|}{\textbf{This Study}} & \textbf{Transition} & \textbf{Changala et al.} & \multicolumn{1}{l|}{\textbf{This Study}} & \textbf{Transition} & \textbf{Changala et al.} & \textbf{This Study} \\ \hline
\textbf{R(19)} & 2.88283 & 3.03821 & \textbf{R(5)} & 0.09044 & 0.06712 & \textbf{P(9)} & 4.09076 & 3.93378 \\
\textbf{R(18)} & 2.41605 & 2.55795 & \textbf{R(4)} & 0.18326 & 0.14975 & \textbf{P(10)} & 4.37871 & 4.21857 \\
\textbf{R(17)} & 1.98722 & 2.11561 & \textbf{R(3)} & 0.30131 & 0.25858 & \textbf{P(11)} & 4.65704 & 4.49416 \\
\textbf{R(16)} & 1.59804 & 1.71294 & \textbf{R(2)} & 0.43506 & 0.38462 & \textbf{P(12)} & 4.9231 & 4.75791 \\
\textbf{R(15)} & 1.24991 & 1.35135 & \textbf{R(1)} & 0.56372 & 0.50836 & \textbf{P(13)} & 5.17457 & 5.00751 \\
\textbf{R(14)} & 0.94394 & 1.03198 & \textbf{R(0)} & 0.61818 & 0.56547 & \textbf{P(14)} & 5.40944 & 5.24094 \\
\textbf{R(13)} & 0.68091 & 0.75564 & \textbf{P(1)} & 2.87367 & 2.67911 & \textbf{P(15)} & 5.62596 & 5.45643 \\
\textbf{R(12)} & 0.46122 & 0.52277 & \textbf{P(2)} & 2.28988 & 2.14908 & \textbf{P(16)} & 5.82263 & 5.65251 \\
\textbf{R(11)} & 0.28493 & 0.33346 & \textbf{P(3)} & 2.41302 & 2.27697 & \textbf{P(17)} & 5.99824 & 5.82790 \\
\textbf{R(10)} & 0.15172 & 0.18743 & \textbf{P(4)} & 2.64446 & 2.50664 & \textbf{P(18)} & 6.15178 & 5.98161 \\
\textbf{R(9)} & 0.06085 & 0.08400 & \textbf{P(5)} & 2.91483 & 2.77344 & \textbf{P(19)} & 6.2825 & 6.11286 \\
\textbf{R(8)} & 0.01117 & 0.02207 & \textbf{P(6)} & 3.20295 & 3.05749 & \textbf{P(20)} & 6.38985 & 6.22109 \\
\textbf{R(7)} & 0.00108 & 0.00009 & \textbf{P(7)} & 3.49882 & 3.34926 &  &  &  \\
\textbf{R(6)} & 0.02845 & 0.01601 & \textbf{P(8)} & 3.79625 & 3.64282 &  &  &  \\ \hline
\end{tabular}%
}
\end{table}

\section{Discussion \& Conclusions}

A rovibronic line list has been calculated for the X\,$^{1}\Sigma^{+}$ and A\,$^{1}\Pi$ states of $^{12}$CH$^{+}$ and $^{13}$CH$^{+}$ for the first time; the PYT line lists provide energy levels with uncertainties, transition frequencies and Einstein $A$ coefficients.
The states, transition and partition function files for the PYT line lists can be downloaded from \href{www.exomol.com}{www.exomol.com} or zenodo.

The line list is used to generate absorption and emission spectra, as well as partition functions, which help to visualise the line list and provide a comparison to previous studies.
The $^{12}$CH$^{+}$ line list is naturally very accurate for $v=0-3$ of both states where
empirical energy levels are comprehensive. The calculated levels, making up the
line list outside the realm of observational data, are more uncertain but our analysis
 shows that
the calculations reproduce experimental line frequencies well (to $\lesssim
0.01$ cm$^{-1}$) up to at least $J = 12$ for X\,$^{1}\Sigma^{+}$ and
A\,$^{1}\Pi(f)$ states. Residuals in A\,$^{1}\Pi(e)$ states are expectedly
larger than their $f$ counterparts since their energies were calculated with
an approximate treatment of $\Lambda$-doubling. These show still accuracies to at least
$\sim0.01$ cm$^{-1}$ for $J\leq 9$; however, the lack of  observational data
for $v\geq 4$ means these data should be treated
with more caution.
In general, agreement is better in the lower energy X\,$^{1}\Sigma^{+}$, as
expected, and in both states residuals increase with greater $v$ and exhibit
an apparent $J^{2}$ dependence at high $J$. Such levels are high in energy
(for example the lowest A\,$^{1}\Pi$ level is over 23\,600 cm$^{-1}$ above
that of X\,$^{1}\Sigma^{+}$) meaning that the discrepancy is likely due, in
part, to interaction with more highly excited electronic states that has not
been considered here, such as spin-orbit coupling, which becomes increasingly
important at higher energies and may perturb energy levels. Note that, where
residuals are relatively large, the MARVEL (obs.) data is also more
uncertain, since these high-energy levels are typically determined by just one
observed transition (usually multiple are available) yielding uncertainties of
the order $1-3$ cm$^{-1}$ for the few $v\geq 4$ empirical levels in both
X\,$^{1}\Sigma^{+}$ and A\,$^{1}\Pi$ states.

The PYT line lists for $^{12}$C$^1$H$^{+}$ and $^{13}$C$^1$H$^{+}$ comprehensively
characterises well above and beyond the domain of previous observation.
Identification of future avenues of research, dependent on later observation,
have also been presented. The accuracy and completeness of the line list open up
its potential application to future high-resolution studies and any hot
observations of CH$^{+}$, as well as the investigation of a range of
astronomical environments, including the complex chemistry of interstellar
clouds.

\section*{Acknowledgments}
 This work was  supported by the European Research Council
(ERC) under the European Union’s Horizon 2020 research and innovation programme
through Advance Grant number 883830 and the STFC Projects No. ST/M001334/1 and ST/R000476/1.
We thank Nike Dattani for providing the copy of LEVEL which we adapted for this study.

\section*{Data Availability}

The states, transition and partition function files for the $^{12}$CH$^+$ and  $^{13}$CH$^+$ line lists can
be downloaded from
\url{www.exomol.com} and zenodo. Inputs for LEVEL, and the $^{12}$CH$^+$ MARVEL energies and transitions files
are provided as supplementary data to this article. ExoCross and our adapted version of LEVEL are available at
\href{github.org/exomol}{github.org/exomol}.


\label{lastpage}
\end{document}